\documentclass[%
superscriptaddress,
 amsmath,amssymb,
 aps,
 prd,
 twocolumn,
floatfix]{revtex4}

\usepackage{graphicx}
\usepackage{amssymb}
\usepackage{amsmath}
\usepackage{hyperref}
\usepackage{color}
\usepackage{url}

\usepackage{dcolumn}
\usepackage{bm}

\newcommand{\beq}{\begin{equation}}
\newcommand{\beqa}{\begin{eqnarray}}
\newcommand{\eeq}{\end{equation}}
\newcommand{\eeqa}{\end{eqnarray}}

\newcommand{\bv}{{\bf v}}

\newcommand{\simgt}{\lower.5ex\hbox{$\; \buildrel > \over \sim \;$}}
\newcommand{\simlt}{\lower.5ex\hbox{$\; \buildrel < \over \sim \;$}}

\def\avrg#1{\left\langle #1 \right\rangle}

\begin{document}


\title{Constraints on Earth-mass primordial black holes from OGLE 5-year microlensing events}

\author{Hiroko~Niikura}
  \email[]{niikura@hep.phys.s.u-tokyo.ac.jp}
\affiliation{%
 Physics Department,
The University of Tokyo, Bunkyo, Tokyo 113-0031, Japan
}%
\affiliation{%
Kavli Institute for the Physics and Mathematics of the Universe (WPI), The University of Tokyo Institutes for Advanced Study (UTIAS),
The University of Tokyo, 5-1-5 Kashiwanoha, Kashiwa, Chiba, 277-8583, Japan
}
\author{Masahiro Takada}
\email[]{masahiro.takada@ipmu.jp}
\affiliation{%
Kavli Institute for the Physics and Mathematics of the Universe (WPI), The University of Tokyo Institutes for Advanced Study (UTIAS),
The University of Tokyo, 5-1-5 Kashiwanoha, Kashiwa, Chiba, 277-8583, Japan
}%
\author{Shuichiro~Yokoyama}
\affiliation{
	Division of Particle and Astrophysical Science, Graduate School of Science, Nagoya
University, Nagoya 464-8602, Aichi, Japan
}
\affiliation{%
Kavli Institute for the Physics and Mathematics of the Universe (WPI), The University of Tokyo Institutes for Advanced Study (UTIAS),
The University of Tokyo, 5-1-5 Kashiwanoha, Kashiwa, Chiba, 277-8583, Japan
}%
\author{Takahiro~Sumi}
\affiliation{
Department of Earth and Space Science, Graduate School of Science, Osaka University, 1-1 Machikaneyama, Toyonaka, Osaka 560-0043, Japan
}
\author{Shogo~Masaki}
\affiliation{
	Department of Mechanical Engineering, National Institute of Technology, Suzuka College, Suzuka, Mie, 510-0294, Japan
}

\date{\today}

\begin{abstract}
We constrain the abundance of primordial black holes (PBH) using 
2622 microlensing events obtained from 5-years observations of stars in the Galactic bulge by 
the Optical Gravitational Lensing Experiment (OGLE).
The majority of microlensing events 
display a single or at least continuous population  that has 
a peak around the light curve timescale $t_{\rm E}\simeq 20~{\rm days}$ and a wide distribution over the range  $t_{\rm E}\simeq [1, 300]~{\rm days}$, while the data also indicates a second population of 6 ultrashort-timescale events 
in $t_{\rm E}\simeq [0.1,0.3]~{\rm days}$, which are advocated to be due to free-floating planets. 
We confirm that the main population of OGLE events can be well modeled by microlensing due to brown dwarfs, main sequence 
stars and stellar remnants (white dwarfs and neutron stars) in 
the standard Galactic bulge and disk models for their spatial and velocity distributions. 
Using the dark matter (DM) model for the Milky Way (MW) halo relative to the Galactic bulge/disk models,  
we obtain the tightest upper bound on the PBH abundance in the mass range $M_{\rm PBH}\simeq[10^{-6},10^{-3}]M_\odot$ (Earth-Jupiter mass range),  
{\it if} we employ 
``null hypothesis'' that the OGLE data does not contain any 
PBH microlensing event. More interestingly, we also show that Earth-mass PBHs can well reproduce the 6 ultrashort-timescale events, without the need of free-floating planets, if the mass fraction of PBH to DM  is at a per cent level, which is consistent with other constraints such as the microlensing search for Andromeda galaxy (M31) and the longer timescale OGLE events. Our result gives a hint of 
PBH existence, and can be confirmed or falsified by  microlensing search for stars in M31, because M31 is towards the MW halo direction and should therefore contain a much less number of free-floating planets, even if exist, than the direction to the MW center.
\end{abstract}


\maketitle

\section{Introduction}

Dark matter (DM) is one of the most essential ingredients in the standard model of cosmic hierarchical structure formation scenario
\cite[e.g.,][]{2005Natur.435..629S}. However, the nature of DM is not yet known and one of the most important, unresolved problems in physics and astronomy. Unknown stable elementary particle(s) beyond the Standard Model of Particle Physics, so-called Weakly Interacting Massive Particle(s) (WIMP), have been thought of as a viable candidate of DM, but has yet to be detected either in direct experiments, collider experiments or indirect searches 
\citep[e.g.,][for a review]{Jungmanetal:96,Arina:2018zcq,Hooper:2018kfv}. 
Primordial black holes (PBH), first proposed 
in pioneer works \cite{Hawking:71,CarrHawking:74,Carr:75},  are alternative viable candidate of DM \citep[also see][for a review]{Carretal:16}. PBHs can be formed by gravitational collapse of Hubble patch in the early universe if the patch has a large primordial overdensity of $\delta\sim O(0.1)$. Various works have proposed a mechanism to generate such a large overdensity such as an inflation model \citep[e.g.,][]{Inomataetal:16,2017PhRvL.119c1103C,2018PhRvD..98h3513C}, producing PBHs with a target mass scale and a target abundance. 
Furthermore, there is a renewed interest in PBH scenario because of recent claims 
\citep{Sasakietal:16,Birdetal:16,2018CQGra..35f3001S,2017PhRvD..96l3523A,2016PTEP.2016i3E01N} \citep[also see][]{1998PhRvD..58f3003I}
that PBHs can be progenitors of 
binary black holes 
whose gravitational wave events have been detected by the LIGO/Virgo experiments \citep{LIGO:16,PhysRevLett.119.141101}.

There are various attempts at constraining PBHs over almost twenty orders of magnitudes in their mass scales; gamma-ray background from PBH evaporation \citep{Carretal:10}, femtolensing of
gamma-ray bursts \citep{Gould:92} (although Ref.~\cite{2018JCAP...12..005K} recently pointed out that a finite-source size effect 
of the gamma-ray burst progenitor significantly relaxes or even removes the constraint), supernovae of white dwarfs triggered 
by PBH \citep{2015PhRvD..92f3007G},
PBH capture by neutron stars \citep{Capelaetal:13b,2017PhRvL.119f1101F}, microlensing constraints \citep{Alcocketal:00,EROS:07,Griestetal:14,2017arXiv170102151N}, 
caustics-network lensing in the galaxy cluster region \citep{2018NatAs...2..334K,2018ApJ...857...25D,2018PhRvD..97b3518O}, $X$-ray background from gas accretion on PBH 
\citep{2017JCAP...10..034I}, and the effect of PBH gas accretion on the cosmic microwave background optical depth 
\citep{Ricottietal:08} (see Refs.~\citep{Ali-HaimoudKamionkowski:17,2017JCAP...05..017A} for the revisited calculations), 
the effect of PBH on pulsar timing array observation \cite{2017PhRvD..95b3002S} and the effect on type-Ia supernova observation
\citep{2018PhRvL.121n1101Z}. Except for a mass window of $M_{\rm PBH}\simeq [10^{-16},10^{-11}]M_\odot$, these constraints rule out a scenario that PBHs constitute the dominant fraction of DM if PBH has a narrow mass spectrum. 

Microlensing is the most robust, powerful tool among various methods to probe a compact, macroscopic DM in the Milky Way (MW) halo region \citep{Paczynski:86,Griestetal:91}, 
because lensing is a gravitational effect and can directly probe mass (gravity strength) of a lensing object irrespective of whether a lensing object is visible or not. The Optical Gravitational Lensing Experiment (OGLE \footnote{\url{http://ogle.astrouw.edu.pl}}) collaboration
\citep{1994ApJ...426L..69U,2015AcA....65....1U} has been making invaluable long-term efforts, more than a decade, to make 
monitoring observations of million stars in the Galactic bulge fields. The OGLE team has been finding more than two thousands of microlensing events and obtained various constraints on exoplanetary systems, brown dwarfs, 
low-mass stars as well as presented even an indication of free-floating planets 
in inter-stellar space \citep{2017Natur.548..183M} \citep[also see][for the similar constraints from the MOA microlensing experiments]{Sumietal:03,Sumietal:11}. 

In this paper we use the 5-years OGLE data containing 2622 microlensing events in \citet{2017Natur.548..183M} to constrain the PBH abundance. Interestingly the OGLE data indicates 6 ultrashort-timescale microlensing events that have their light curve timescales of 
$t_{\rm E}\simeq [0.1,0.3]~{\rm days}$ \citep[also see][for the new candidates]{2018arXiv181100441M}, which is a distinct population from the majority of OGLE events. The ultrashort-timescale events indicate Earth-mass ``unbounded'' (wide-orbit or free-floating) planets \citep{Sumietal:11,2017Natur.548..183M}. However, the origin of such free-floating planets is poorly understood because it involves complicated physics of star formation, planetary system formation and interaction of planetary system with other stars/planets. Hence we pay a particular attention to a possibility of whether PBHs can give an alternative explanation of the ultrashort-timescale events. For this purpose, we first study the standard Galactic bulge and disk models to estimate event rates of microlensing events due to astrophysical objects including brown dwarfs, main sequence stars and stellar remnants (white dwarfs, neutron stars and astrophysical black holes) following the pioneer work in \citet{1995ApJ...447...53H} \citep[also see][]{1996ApJ...467..540H}. After comparing the model predictions of astrophysical objects with  
the OGLE data including the calibration factor that takes into account observational effects, we use the OGLE data to constrain the PBH abundance using the standard MW halo model for the spatial and velocity distributions of DM (therefore PBHs). In doing this we employ two working hypotheses. First, we employ ``null hypothesis'' that there is no PBH microlensing in the OGLE data and then derive an upper bound on the PBH abundance. Second, we employ the assumption that the 6 ultrashort-timescale events are due to PBHs and derive an allowed region of PBHs in the mass and abundance parameter space. To obtain the results, we properly use the likelihood function of OGLE events assuming the Poisson uncertainty in the counts of microlensing events.

The structure of this paper is as follows. In
Section~\ref{sec:microlensing}, we briefly review basics of microlensing and give equations relevant for the event rate calculations. 
In Section~\ref{sec:density_velocity} we review the standard models for the Galactic disk and bulge describing the spatial and velocity distributions for 
brown dwarfs, stars and stellar remnants as the constitutions, and also describe the MW halo model for the distributions
of DM, i.e. PBH in our study. 
In Section~\ref{sec:results}, we give the main results of this paper, after reviewing the OGLE data; the upper bound on the PBH abundance and a possible implication of Earth-mass PBHs. 
We will then give conclusion and discussion in Section~\ref{sec:discussion}.

\section{Microlensing for bulge stars}
\label{sec:microlensing}

\subsection{Microlensing basics}
\label{subsec:microlensing_basics}

When a source star and a lensing PBH are almost perfectly aligned along the line-of-sight direction of an observer, the star is multiply imaged due to strong lensing \citep{Paczynski:86} \citep[also see][for a review]{2006glsw.conf.....M,2017grle.book.....D}.
In case these multiple images are unresolved, the flux from the star appears magnified. The light curve of such a microlensing magnification event is given by
\begin{align}
& A(t)=\frac{u^2+2}{u\sqrt{u^2+4}},
\end{align}
where $u(t)$ is the separation between the source star and the lens at an observation epoch $t$, and is given by
\begin{align}
u(t)=\sqrt{\beta^2+
\frac{(t-t_0)^2}{t_{\rm E}^2}}.
\end{align}
Here $\beta$ is the dimensional-less impact parameter, and $t_0$ is the time where the source and the lens become closest in separation.
Throughout this paper we 
use a crossing time of the Einstein radius, 
denoted as $t_{{\rm E}}$, to characterize 
a timescale of microlensing light curve: 
%
\begin{align}
t_{\rm E}
\equiv \frac{R_{\rm E}}{v}=\frac{\sqrt{4GM d_{\rm l}d_{\rm ls}/d_{\rm s}}}{cv},
\label{eq:omega}
\end{align}
where $R_{\rm E}$ is the Einstein radius, $M$ is the mass of lensing object (we assume a point mass for lens throughout
this paper), $v$ is the (total) relative velocity on the two-dimensional 
plane perpendicular to the line-of-sight direction (see below), 
$d_{\rm l}$ and $d_{\rm s}$ are distances to 
the lens and source, respectively, and $d_{\rm ls}$ is the distance between lens and source ($d_{\rm ls}\equiv 
d_{\rm s}-d_{\rm l}$).

If we plug typical values of the physical quantities into Eq.~(\ref{eq:omega}), we can find a typical timescale of the microlensing 
light curve as
\begin{align}
&t_{\rm E}\simeq 44~{\rm days}\left(\frac{M}{M_\odot}\right)^{1/2}
\left(\frac{d_{\rm l}d_{\rm ls}/d_{\rm s}}{4~{\rm kpc}}\right)^{1/2}\left(\frac{v}{220~{\rm km/s}}\right)^{-1}.
\label{eq:t_E}
\end{align}
%
This equation shows that a lighter-mass lens causes a shorter timescale light curve.  
As we will show below, the 5-year OGLE data gives a sufficient sampling of microlensing light curves over the range of 
timescales, $[10^{-1},300]~$days, roughly corresponding to lenses of mass scales, $[10^{-6},10]~M_\odot$. In reality,
since lensing objects 
have a velocity distribution, it 
causes a distribution of the light curve timescales even if lensing objects are in a narrow mass bin,
 which we need to properly take into account. 


\subsection{Microlensing optical depth and event rate for a star in the Galactic bulge}
\label{subsec:optical_depth}

\subsubsection{Definition of microlensing optical depth and event rate}

Here we define the optical depth and event rate of microlensing for a {\em single} star in the Galactic bulge region. The optical depth 
is defined as the probability for a source star to be inside the Einstein radius of a foreground lensing object on the sky at a certain {\em moment}. This corresponds to the probability for the lensing magnification to be greater than $A\ge 1.34$. The total optical depth 
due to lensing objects in the bulge and disk regions as well as due to PBHs in the MW halo region is formally expressed as
\begin{align}
& \tau \equiv \tau_{\rm b}+\tau_{\rm d}+\tau_{\rm PBH}.
\end{align}
Hereafter we employ abbreviations: ``b'' for ``bulge'' and ``d'' for ``disk'', respectively, and 
we ignore a multiple lensing case for a single star (this is a good approximation given the low optical depth as we show below). 
For a lensing object in the bulge and disk regions, we consider brown dwarfs and 
stellar components, where the latter includes main sequence stars and stellar 
remnants (white dwarfs, neutron stars and astrophysical black holes), as we will explain in detail later.

The differential event rate of microlensing is defined as the frequency of microlensing events of a given 
light curve timescale (denoted as $t_{\rm E}$) for a single source star per unit observational time ($t_{\rm obs}$): 
\begin{align}
& \frac{\mathrm{d}\Gamma_a}{\mathrm{d}t_{\rm E}}\equiv \frac{\mathrm{d}^2\tau_a}{\mathrm{d}t_{\rm obs}\mathrm{d}{t}_{\rm E}},
\end{align}
where the subscript $a={\rm bulge}$, disk or PBH, respectively. 

\subsubsection{Coordinate system}
\label{sec:coordinate}

It would be useful to explicitly define the coordinate system we employ in the following calculations. For the rectangular coordinate system, 
denoted as $(x,y,z)$, we choose the Galactic center as 
the coordinate origin.
Without loss of generality, we can take the $x$-direction to be 
along the direction connecting the Galactic center and the Earth position (an observer's position). 
We assume that 
the Earth 
is located at the position, $(x,y,z)_{\oplus}=(8~{\rm kpc}, 0, 0)$, i.e. 8~kpc in distance
from the Galactic center. 
Furthermore, we take the $y$-direction to be along the 
Earth's rotation direction in the Galactic disk plane, 
and the $z$-direction to be in the direction perpendicular 
to the disk plane.

In this paper we consider the microlensing datasets obtained from the 5-years OGLE survey \citep{2015AcA....65....1U}\footnote{The OGLE-IV 
fields can be found from \url{http://ogle.astrouw.edu.pl/sky/ogle4-BLG/}.}. In the Galactic coordinates, the OGLE fields are located in the range of 
$-15^{\circ}\simlt b\simlt 15^{\circ}$ and $-20^{\circ}\simlt l \simlt 20^\circ$. Throughout this paper we simply assume that the OGLE field is in the
direction to 
the field BLG505 with $(b,l)=(-2^\circ\!\!.389,1^\circ\!\!.0879)$, 
which has the largest number of background stars among the OGLE fields. 
We believe that this approximation is valid because our results are based on 
relative contributions of microlensing due to stellar components in the bulge and disk regions compared 
to microlensing due to PBHs in the MW halo region. 

\subsubsection{Bulge lens}
\label{sec:bulge_gamma}

First we consider a microlensing that both lens and source are in the Galactic bulge region. 
The {\em average} optical depth of microlensing due to the $i$-th stellar component 
for a single source star 
is given by
\begin{align}
\tau_{{\rm b}} &\equiv \frac{1}{N_{\rm s}}
\int_{d_{\rm s,min}}^{d_{\rm s,max}}\!\mathrm{d}d_{\rm s}~n_{\rm s}(d_{\rm s})
\sum_{i}
\int_{d_{\rm s,min}}^{d_{\rm s}}\!\mathrm{d}d_{\rm l}~\frac{\rho_{{\rm b},i}(d_{\rm l})}{M_i}\pi R_{\rm E}^2(M_i)
\nonumber\\
&=\frac{4\pi G}{c^2N_{\rm s}}\int_{d_{\rm s,min}}^{d_{\rm s,max}}\!\mathrm{d}d_{\rm s}~n_{\rm s}(d_{\rm s})
\sum_i \int_{d_{\rm s,min}}^{d_{\rm s}}\!\mathrm{d}d_{\rm l}~\rho_{{\rm b},i}(d_{\rm l})D,
\label{eq:opt_bulge}
\end{align}
where the integration is along the line-of-sight direction (see below),
$D\equiv d_{\rm l}d_{\rm ls}/d_{\rm s}$, and 
the index $i$ stands for the $i$-th stellar component as a lensing object for which 
we will consider brown dwarfs, main sequence stars, white dwarfs, neutron stars and astrophysical black holes  (see below).
$N_{\rm s}$ is the surface number density of source stars defined by a line-of-sight 
integration of the three-dimensional number density distribution of source stars, $n_{{\rm s}}$, as
\begin{align}
N_{\rm s}\equiv \int_{d_{\rm s, min}}^{d_{\rm s, max}}\!\mathrm{d}d_{\rm s}~n_{\rm s}(d_{\rm s}).
\end{align}
The function $\rho_{{\rm b},i}(d_{\rm l})$ is the mass density profile for the $i$-th stellar component. 
$d_{\rm s, min}$ and $d_{\rm s, max}$ are the maximum and minimum distances to the boundary of the bulge region from an observer's position. Throughout this paper
we employ $d_{\rm s, min}=4~{\rm kpc}$ and $d_{\rm s, max}=12~{\rm kpc}$; that is, we assume that the bulge has a size of $4~{\rm kpc}$ around the center ($d_{\rm s}=8~{\rm kpc}$) in depth from an observer. 
The integration over $d_{{\rm s}}$ or $d_{{\rm l}}$ in Eq.~(\ref{eq:opt_bulge}) is along the line-of-sight direction of an observer towards 
the source star in the direction ($y,z$). As long as the lens distribution $\rho_{{\rm b},i}$
and the source star distribution are given,  
the line-of-sight integration is straightforward to perform as we will show later.

Now we consider the event rate of microlensing.  
To do this, we start from the geometry and variables defined in Fig.~4 of 
\citet{Griestetal:91} \citep[also see Figure~7 of][]{2017arXiv170102151N}, which defines the differential event rate of a lensing object entering a volume element 
along the line-of-sight where the lens causes a microlensing with magnification above a certain threshold value:
\begin{align}
\mathrm{d}\Gamma_{{\rm b}}&=\sum_i\frac{\rho_{{\rm b},i}}{M_{i}}R_{\rm E} v_\perp^2\cos\theta 
\mathrm{d}d_{\rm l}\mathrm{d}\alpha
f_{{\rm b},i}({\bf v}_\perp,v_\parallel)\mathrm{d}{v}_\perp\mathrm{d}\theta\mathrm{d}v_\parallel~,
\end{align}
where $f_{{\rm b},i}(\bv_\perp,v_\parallel)$ is the velocity distribution of the $i$-th stellar component,
 defined so as to satisfy 
the normalization condition $\int\!\!\mathrm{d}^2\bv{\perp}\int\!\!\mathrm{d}v_{\parallel}~ f(\bv_\perp,v_\parallel)=1$; $\bv{\perp}$
is the perpendicular components of relative velocity between an observer, lens and source star (see below), defined as $\bv_{\perp}=v_{\perp}(\cos\theta,\sin\theta)$; $\alpha$ is the azimuthal angle in the two-dimensional ($y,z$)-plane, defined as $(y,z)=\sqrt{y^2+z^2}(\cos\alpha,\sin\alpha)$;
$\theta$ is the angle between the line connecting the source and the lens center and the direction of the transverse velocity $\bv_{\perp}$. 
In this paper we define the microlensing ``event'' if the lensing magnification is greater than a threshold magnification, $A>
A(R_{\rm E})=1.34$, 
which is satisfied if the separation between lens and source is closer than the threshold separation, $b\le R_{{\rm E}}$.
The parameters vary in the range of $\theta\in [-\pi/2,\pi/2]$, $\alpha\in[0,2\pi]$, and $v_\perp\in [0,\infty)$.

We assume that the velocity distribution can be simplified as
\begin{align}
f_{\rm b}(\bv_\perp,v_\parallel)=f_{\rm b}(\bv_\perp)f_{\rm b}(v_\parallel).
\end{align}
As we discussed, for a characteristic timescale of microlensing light curve, we employ a crossing time scale of the Einstein ring, defined 
as $t_{\rm E}=2R_{\rm E}\cos\theta/v_\perp$. 
This simplification is not critical for
the following discussion because we study the PBH microlensing contribution relative to those due to the stellar components in the disk and bulge regions. In this case, the microlensing event rate due 
to the $i$-th stellar components is given as
\begin{align}
\frac{\mathrm{d}\Gamma_{{\rm b}}}{\mathrm{d}t_{\rm E}}
&=\frac{2\pi}{N_{\rm s}}\int^{d_{\rm s, max}}_{d_{\rm s, min}}\!\!\mathrm{d}d_{\rm s}~n_{\rm s}(d_{\rm s})
\sum_i\int^{d_{\rm s}}_{d_{\rm s,min}}\!\!\mathrm{d}d_{\rm l}
\frac{\rho_{{\rm b},i}(d_{\rm l})}{M_i}R_{\rm E}(d_{\rm s},d_{\rm l})\nonumber\\
&\hspace{2em}\times \int_0^\infty\!\!\mathrm{d}v_\perp\int_{-\pi/2}^{\pi/2}\!\!\mathrm{d}\theta~v_\perp^2
\cos\theta f_{{\rm b},i}(v_\perp,\theta)\nonumber\\
&\hspace{2em}\times \delta_D\!\left(t_{\rm E}-\frac{2R_{\rm E}\cos\theta}{v_\perp}\right).
\end{align}
Using the Dirac delta function identity
\begin{align}
\delta_D\!\left(t_{\rm E}-\frac{2R_{\rm E}\cos\theta}{v_\perp}\right)=\delta_D\!\left(v_\perp-\frac{2R_{\rm E}\cos\theta}{t_{\rm E}}\right)\frac{v_\perp^2}{2R_{\rm E}\cos\theta},
\end{align}
the above equation is simplified as
\begin{align}
\frac{\mathrm{d}\Gamma_{{\rm b}}}{\mathrm{d}t_{\rm E}}
&= \frac{\pi}{N_{\rm s}}\int^{d_{\rm s, max}}_{d_{\rm s, min}}\!\!\mathrm{d}d_{\rm s}~n_{\rm s}(d_{\rm s})
\nonumber\\
&\hspace{2em} \times
\sum_i 
\int^{d_{\rm s}}_{d_{\rm s,min}}\!\!\mathrm{d}d_{\rm l}
\frac{\rho_{{\rm b},i}(d_{\rm l})}{M_i}\int_{-\pi/2}^{\pi/2}\!\!\mathrm{d}\theta~v_\perp^4 
f_{{\rm b},i}(v_\perp,\theta),
\label{eq:gamma_b}
\end{align}
where $v_{\perp}=2R_{{\rm E}}\cos\theta/t_{{\rm E}}$. With this condition, 
the tangential velocity $v_\perp$ depends on integration variables, $d_{\rm s}, d_{\rm l}$, and $\theta$ via $R_{\rm E}=R_{\rm E}(d_{\rm l},d_{\rm s})$.

\subsubsection{Disk lens}

Next we consider an event rate for microlesning due to stellar components in the disk region for a single source star in the bulge region.
The calculation is very similar to the case for bulge lens in the preceding section. 
In this case we employ a single source plane approximation for simplicity; that is, 
we assume that all source stars are at distance of $8~$kpc, the Galactic center. 
Under this assumption, the optical depth is given by 
\begin{align}
\tau_{{\rm d}}=\frac{4\pi G}{c^2}\int_0^{\bar{d}_{\rm s}}\!\!\mathrm{d}d_{\rm l}~\sum_i \rho_{{\rm d},i}(d_{\rm l})D,
\end{align}
where $\rho_{{\rm d},i}(d_{\rm l})$ is the mass density distribution of the $i$-th stellar component, 
$D=d_{{\rm l}}d_{{\rm ls}}/\bar{d}_{{\rm s}}$,  $\bar{d}_{{\rm s}}$ is the mean distance to source stars, i.e.
$\bar{d}_{\rm s}=8~{\rm kpc}$, and $d_{{\rm ls}}=\bar{d}_{{\rm s}}-d_{{\rm l}}$.

Similarly, the event rate is
\begin{align}
\frac{\mathrm{d}\Gamma_{{\rm d}}}{\mathrm{d}t_{\rm E}}
=\pi\sum_i\int^{\bar{d}_{\rm s}}_{0}\!\!\mathrm{d}d_{\rm l}
\frac{\rho_{{\rm d},i}(d_{\rm l})}{M_i}\int_{-\pi/2}^{\pi/2}\!\!\mathrm{d}\theta~v_\perp^4 
f_{{\rm d},i}(v_\perp,\theta),
\label{eq:gamma_d}
\end{align}
where $v_{\perp}=2R_{{\rm E}}\cos\theta/t_{{\rm E}}$, $f_{{\rm d},i}(\bv_\perp)=f_{{{\rm d}},i}(v_\perp,\theta)$ is the velocity distribution for velocity components perpendicular to the 
ling-of-sight direction for
the $i$-th stellar component in the disk region.

\subsubsection{PBH lens}

Now we consider a scenario that PBHs constitutes 
some mass fraction of DM in the MW halo region. We call ``PBHs in the halo region'' because PBHs are distributed from the Galactic center through the outer halo region due to 
the large velocity dispersion. When a lensing PBH happens to pass across a source star in the bulge on the sky, it could cause microlensing effect on the source star. 
Throughout this paper we consider a monochromatic mass distribution for PBHs. Similarly to the disk microlensing,
the optical depth of microlensing due to PBHs is 
\begin{align}
&\tau_{\rm PBH}=\frac{4\pi G}{c^2}\int_0^{\bar{d}_{\rm s}}\!\!\mathrm{d}d_{\rm l}~ {\rho}_{\rm DM}(d_{\rm l})D,
\end{align}
where $\rho_{{\rm DM}}(d_{\rm l})$ is the dark matter distribution. If PBHs consist only some partial mass fraction of DM in the MW region, denoted as $f_{\rm PBH}\equiv \Omega_{\rm PBH}/\Omega_{\rm DM}$, 
we replace $\rho_{\rm DM}$ in the above and following equations with $f_{\rm PBH}\rho_{\rm DM}$.

Similarly, the event rate of microlensing due to PBHs for a single source star in the bulge is 
\begin{align}
\frac{\mathrm{d}\Gamma_{{\rm PBH}}}{\mathrm{d}t_{\rm E}}
=\pi\int^{\bar{d}_{\rm s}}_{0}\!\!\mathrm{d}d_{\rm l}
\frac{\rho_{{\rm DM}}(d_{\rm l})}{M_{\rm PBH}}\int_{-\pi/2}^{\pi/2}\!\!\mathrm{d}\theta~v_\perp^4 
f_{{\rm DM}}(v_\perp,\theta),
\label{eq:gamma_pbh}
\end{align}
where $f_{{\rm DM}}(\bv_\perp)$ is the velocity distribution of PBHs. 

\section{Models of Galactic disk and bulge and Milky Way dark matter}
\label{sec:density_velocity}

As we described, 
once we give the density and velocity distributions for stellar components in the MW bulge and 
disk regions as well as those for PBHs (equivalently DM) in the halo region, we can compute the 
event rates of microlensing for a star in the bulge region.
In this subsection, we briefly review the standard model for the MW bulge and disk following 
\citet{1995ApJ...447...53H}.  Then we describe our model for the density and velocity distributions for PBHs 
in the MW region.

\subsection{The mass density distribution}
\label{sec:rho}

\begin{table*}
\begin{center}
\begin{tabular}{l|lll}\hline\hline
lens & mass density profile: $\rho$ [$M_\odot\mathrm{pc}^{-3}$]\hspace{5em} & $\tau$ [$10^{-6}$]\hspace{2em} &velocity profile: $(\mu,\sigma)$~[km/s] \\
\hline
bulge & $1.04\times10^6\left(\frac{s}{0.482~{\rm pc}}\right)^{-1.85}, \text{($s<938$pc)}$ 
& 1.07& $f_y: \left\{-220(1-\alpha),\sqrt{1+\alpha^2}100\right\}$ \\
& $3.53~ K_0\!\left(\frac{s}{667~{\rm pc}}\right),  \text{($s\ge938$pc)}$ 
& & $f_z: \left\{0,\sqrt{1+\alpha^2}100\right\}$  \\ \hline
disk & $0.06\times \exp\left[-\left\{\frac{R-8000}{3500}+\frac{z}{325}\right\}\right]$
&1.03& $f_y: \left\{220\alpha,\sqrt{(\kappa\delta+30)^2+(100\alpha)^2}\right\}$ \\
& 
&& $f_z: \left\{0,\sqrt{(\lambda\delta+30)^2+(100\alpha)^2}\right\}$\\ \hline
PBH & $4.88\times10^{-3}f_{\rm PBH}\frac{1}{(r/r_s)(1+r/r_s)^2}$ 
&0.18$f_{\rm PBH}$& $f_y: \left\{-220(1-\alpha),\sqrt{\sigma_{\rm DM}^2+(\alpha100)^2}\right\}$ \\
&
&& $f_z: \left\{0,\sqrt{\sigma_{\rm DM}^2+(\alpha100)^2}\right\}$ 
\\
\hline\hline
\end{tabular}
\caption[Summary of the Galactic models] {\label{tab:model_summary}Summary of the Galactic models for the mass and 
velocity distributions for stellar components and PBHs. $\alpha$ is the ratio of distances between lens and source, $\alpha\equiv 
d_{\rm l}/d_{\rm s}$. We employ the coordinate system as defined in Section~\ref{sec:coordinate}, and 
take the Galactic center as the 
coordinate origin. 
The optical depth ($\tau$) is calculated assuming an observation in the direction of $(b,l)=(-2^\circ\!\!.389,1^\circ\!\!.0879)$ which represents the OGLE Galctic bulge fields. 
For PBH case, we assume that PBHs constitute DM in the MW region by mass fraction, $f_{\rm PBH}\equiv \Omega_{\rm PBH}/\Omega_{\rm DM}$, when computing the microlensing optical depth. We assume a Gaussian for the velocity profile, and 
the quantities, $\mu$ and $\sigma$, denote
the mean and dispersion for the Gaussian distribution. For PBH, we employ $\sigma_{\rm DM}=220~{\rm km/s}$ for our fiducial model,
 which is taken from the rotational velocity of Galactic disk (see text). 
\label{tab:rho}
}
\end{center}
\end{table*}
\begin{table*}
\begin{center}
\begin{tabular}{lllll}
\hline\hline
object & parameters in $\mathrm{d}n/\mathrm{d}M$ & mass range $[M_\odot]$ & initial mass range $[M_\odot]$
& $N$\\
\hline 
brown dwarf (BD) & Power-law ($M^{-0.8}$) & [0.01,\,0.08] & $0.01 \leq M \leq 0.08$ &
0.18\\
main-sequence star (MS) & Power-law ($M^{-2}$) & [0.5,\,1.0] & $0.5 \leq M \leq 1.0$ &
1\\
 & Power-law ($M^{-1.3}$) & [0.08\,0.5] & $0.08 \leq M \leq 0.5$ &\\
white dwarf (WD) & Power-law (initially $M^{-2}$) & [0.34,\,2.0] & $1.0 \leq M \leq 8.0$ &
0.15\\
neutron star (NS) & Gaussian ($M_r=1.33, \,\sigma_r=0.12$) & [0.73,\,1.93] & $8.00 \leq M \leq 20.0$ & 
0.013\\
black hole (BH) & Gaussian ($M_r=7.8, \,\sigma_r=1.2$) & [1.8,\,13.8] & $20.0 \leq M \leq 100.0$ & 
0.0068\\ \hline\hline
\end{tabular}
\caption{Summary of the mass spectrum for each of astrophysical objects: brown dwarfs, main-sequence 
stars, and stellar remnants (white dwarfs, neutron stars, and astrophysical black holes) in the standard 
Galactic bulge and disk models. We assume
the Kroupa initial mass function as shown in Fig.~\ref{fig:IMF}, and then assume that 
each massive star
with  initial masses $M\ge 1M_\odot$,   as denoted in the column ``initial mass range'', 
evolved into each stellar remnant. For white dwarf, we assume the relation between initial and end masses as given by 
$M_{\rm WD}=0.339+0.129M_{\rm init}$. The column ``$\mathrm{d}n/\mathrm{d}M$'' denotes parameters of 
the mass spectrum for each object population, while we assume a Gaussian distribution with the mean and width values
for the mass spectrum of 
neutron stars and black holes. The last column ``$N$'' gives the number of each object population relative to that of main-sequence 
stars used in the calculation of microlensing event rate.
\label{tab:massfunc}}
\end{center}
\end{table*}
For the mass density distribution of stellar population in the bulge region, we adopt the model in \citet{1992ApJ...387..181K} 
that describes  the following bar-structured model: 
\begin{align}
&\rho_\mathrm{b}(x,y,z)\nonumber\\
&=
\begin{cases}
1.04\times10^6\left(\frac{s}{0.482~{\rm pc}}\right)^{-1.85}~M_\odot\mathrm{pc^{-3}},
&\text{($s<938$pc)},\\
3.53~ K_0\!\left(\frac{s}{667~{\rm pc}}\right)\,M_\odot\mathrm{pc^{-3}}, &\text{($s\ge938$pc)}, 
\end{cases}
\label{eq:bulgeiso}
\end{align}
where $K_{0}(x)$ is the modified Bessel function, $s$ is the radius from the Galactic center in the elliptical coordinates, 
defined as $s^4\equiv R^4+(z/0.61)^4$ with 
$R\equiv (x^2+y^2)^{1/2}$, and all the coordinate components ($x,y,z,s, R$) are in units of pc. 
As defined in Section~\ref{sec:coordinate}, the coordinate origin is the Galactic center. 
Note that the above profile is continuous at $s=938~{\rm pc}$, and we consider the above profile is for the total 
contribution of visible objects, i.e main sequence stars, as we will describe below. Using the Galactic celestial coordinate variables $(l,b)$, 
a star at the distance $d$ from an observer (the Earth's position) is at the distance from the Galactic center, $r$, given as
\begin{align}
r(d)=\sqrt{R_\oplus^2-2R_\oplus d\cos l \cos b+r^2},
\end{align}
where $r=\sqrt{x^2+y^2+z^2}$, $x=d\cos b\cos l$, $y=d\cos b\sin l$, 
and $z=d\sin b$. This variable transformation between $d$ ($d_{\rm l}$ or $d_{\rm s}$) and ${x,y,z}$ enters into the above 
equations such as Eq.~(\ref{eq:gamma_b}).

For the mass density distribution in the disk region, we employ the model in \citet{1986ARA&A..24..577B}:
\begin{equation}
\rho_\mathrm{d}(R,z)=0.06\times \exp\left[-\left\{\frac{R-8000}{3500}+\frac{z}{325}\right\}\right]~M_\odot\mathrm{pc}^{-3}.
\label{eq:diskiso}
\end{equation}
Note that, as we defined,  $R(=\sqrt{x^2+y^2})$ denotes 
the radial distance in the cylindrical coordinates and $z$ is in the direction perpendicular to the Galactic disk
(variables are in units of pc). 
This model assumes that the disk has an exponential distribution with vertical and radial scale lengths of 325~pc and 3500~pc, respectively. 
Although the mass-to-light ratio of disk stellar population is not well understood, we normalize the above 
density profile to $\rho_{\rm d0}=0.06~M_\odot\mathrm{pc}^{-3}$ at the solar neighborhood ($R=8000~{\rm pc}$).

For the spatial distribution of DM (therefore PBHs) between the Galactic center 
and an observer (the Earth), we assume the Navarro-Frenk-White (NFW) model \citep{NFW97}:
\begin{equation}
\rho_\mathrm{NFW}(r)= \frac{\rho_{c}}{(r/r_{s})(1+r/r_{s})^2},
\label{eq:rho_nfwm}
\end{equation}
where $r_s$ is
the scale radius and $\rho_c$ is the central density parameter. For this model
we assume spherical symmetry for the DM distribution for simplicity. 
In this
paper we adopt the halo model in 
\citet{Klypinetal:02}: $M_{\rm
vir}=10^{12}M_\odot$, $\rho_c=4.88\times 10^{6}~{M_\odot/{\rm kpc}^3}$,
and $r_s=21.5~{\rm kpc}$, taken from Table~2 in the paper. 
The DM profile with these parameters has been
shown to fairly well reproduce the observed rotation curve in the MW. 
However, there might still be a residual uncertainty in the total mass (mostly DM) of MW within a factor of 2
\citep{2018arXiv180810456C}. 

Table~\ref{tab:rho} summarizes models of the mass density profiles for stellar components in the bulge and
disk regions and for dark matter in the MW halo region. The table also gives the optical depth for a single
source star for lenses in the bulge or disk region and for PBHs, respectively. Note that the optical depth 
does not depend on a lens mass as indicated from Eq.~(\ref{eq:opt_bulge}). The table shows that the optical depth due to PBHs is smaller than that of astrophysical objects in the disk or bulge region, by a factor of 5, reflecting that astrophysical objects are more centrally concentrated due to the dissipation processes. 

\subsection{The velocity distribution}
\label{sec:f_v}

A timescale of 
the microlensing light curve (see Eqs.~\ref{eq:gamma_b}, \ref{eq:gamma_d}, and \ref{eq:gamma_pbh})
is determined by a transverse component of the relative velocity for source-lens-observer system on the sky
\citep{Griestetal:91,1995ApJ...447...53H}:
\begin{align}
\bv_\perp &= \bv_{\rm l}-\left(\frac{d_{\rm l}}{d_{\rm s}}\bv_{\rm s}+\frac{d_{\rm ls}}{d_{\rm s}}\bv_{\rm o}\right)\nonumber\\
&= \bv_{\rm l}-\left[\alpha\bv_{\rm s}+(1-\alpha)\bv_{\rm o}\right],
\end{align}
where $\bv_{{\rm l}}$, $\bv_{\rm s}$ and $\bv_{\rm o}$ are the transverse velocities for lens,  source star and an observer, 
respectively, and we have introduced the notation $\alpha\equiv d_{{\rm l}}/d_{{\rm s}}$. As we described in Section~\ref{sec:coordinate}, 
the $x$-direction is along the direction from the observer to the Galactic center (i.e. a source star), which is equivalent to the line-of-sight 
direction, 
the $y$-direction is along
the direction of disk rotation, and the $z$-direction is perpendicular to the line-of-sight direction. Hence we need to model the mean and distribution of the transverse velocity components, $\bv_\perp=(v_y,v_z)$. Hereafter we often omit the subscript ``$\perp$'' 
in $\bv_\perp$ for notational simplicity.

\subsubsection{Bulge lens}
\label{sec:bulge_vel}

First we consider the velocity distribution for the bulge microlensing where both lens and source star are in the bulge region. 
For the velocity distribution, we assume that the stellar components are supported by
an isotropic velocity dispersion, and do not have any rotational velocity component. 
Under these assumptions, the mean of the transverse velocities is
\begin{align}
&\bar{v}_{{\rm b}y} \equiv  \avrg{v_{{\rm l}y}-\left[\alpha v_{{\rm s}y}-(1-\alpha)v_{{\rm o}y}\right]}=
-220(1-\alpha)~\mbox{km/s}, \nonumber\\
&\bar{v}_{{\rm b}z}=0
\end{align}
where we have assumed that an observer is in the rest frame of the rigid body rotation of Galactic disk, has the rotational velocity of 
220~km/s with respect to the Galactic center, and has no mean velocity in the disk height direction. 

The velocity dispersion for the $y$-component of relative velocity can be computed as
\begin{align}
 \sigma_{{\rm b}y}^2 &\equiv \avrg{(v_y)^2}-\avrg{v_y}^2\nonumber\\
&= \avrg{v_{{\rm l}y}^2}+\alpha^2\avrg{v_{{\rm s}y}^2}\nonumber\\
&=(1+\alpha^2) (100~\mbox{km/s})^2,
\end{align}
Here we assumed that the velocity dispersion per component $\sigma_{y}=100$~km/s, and 
assumed that the source and lens have independent random motions; $\avrg{v_{{\rm s}y}v_{{\rm l}y}}=0$.
The velocity dispersion for the velocity $z$-component is 
\begin{align}
\sigma_{{\rm b}z}^2=(1+\alpha^2)(100~{\rm km/s})^2. 
\end{align}

Following \citet{1995ApJ...447...53H}, we assume that the velocity distribution is given by a Gaussian and that 
the velocity distribution for the bulge microlensing is given by
\begin{align}
f_{\rm b}(\bv)=f_{\rm b}(v_y)f_{\rm b}(v_z), 
\end{align}
where 
\begin{align}
&f_{\rm b}(v_y)=\frac{1}{\sqrt{2\pi}\sigma_{{\rm b}y}}
\exp\left[-\frac{\left(v_y-\bar{v}_{{\rm b}y}\right)^2}{2\sigma_{{\rm b}y}^2}\right], \nonumber\\
&f_{\rm b}(v_z)=\frac{1}{\sqrt{2\pi}\sigma_{{\rm b}z}}
\exp\left[-\frac{v_z^2}{2\sigma_{{\rm b}z}^2}\right],
\label{eq:f_b}
\end{align}

\subsubsection{Disk lens}
\label{sec:disk_vel}

Next we consider the velocity distribution for stellar components in the disk region. 
We assume that the stellar 
components have a rigid rotation on average: 
\begin{align}
&\bar{v}_{{\rm d}y}=220\alpha~{\rm km/s}\nonumber\\
&\bar{v}_{{\rm d}z}=0.
\end{align}
For the velocity dispersion, we assume the linear disk velocity dispersion model in Table~1 of \citet{1995ApJ...447...53H}:
\begin{align}
&\sigma_{{\rm d}y}^2=(\kappa\delta+30)^2+(100\alpha)^2~({\rm km/s})^2\nonumber\\
&\sigma_{{\rm d}z}^2=(\lambda\delta+20)^2+(100\alpha)^2~({{\rm km/s}})^2
\end{align}
with 
\begin{align}
&\kappa\equiv 5.625\times 10^{-3}~{\rm km/s/pc},\nonumber \\
&\lambda\equiv 3.75\times 10^{-3}~{\rm km/s/pc},\nonumber \\
&\delta\equiv (8000-x)~{\rm pc}~ , 
\end{align}
where $\kappa$ and $\lambda$ are the velocity dispersion gradient coefficients. 
Hence the velocity distribution functions are given by the similar equations to Eq.~(\ref{eq:f_b}).

\subsubsection{PBH lens}
\label{sec:pbh_vel}

Now we consider a microlensing due to PBHs, acting as DM, in the MW halo region. PBHs are tracers of the MW halo that is much more extended than the bulge size ($\sim 200$~kpc vs. a few kpc in radius). 
The large extent of DM halo reflects the fact that PBHs have a 
larger velocity dispersion than that of bulge stars (100~km/s). 
First we assume that PBHs have isotropic velocity distribution with respect to the halo center for which we assume the Galactic center. Hence the mean relative  velocity for a PBH lens is 
\begin{align}
&\bar{v}_{{\rm PBH}y}=-220(1-\alpha)~{\rm km/s},\nonumber\\ 
&\bar{v}_{{\rm PBH}z}=0.
\end{align}
For a PBH causing microlesning effect on a bulge star, it should be located somewhere
between the Galactic center and the Earth, which is a very inner region compared to the halo size. Such a PBH (more generally dark matter) tends 
to have a large velocity when passing through the central region of DM halo; 
DM tends to have a larger velocity at the closest point to the halo center (i.e. the Galactic center), while a bounded DM should stop at an apocenter point of its orbit, which tends to be around the outer boundary of the halo. Hence we assume
that PBHs causing the microlensing have a large velocity dispersion whose amplitude is similar to the rotation velocity. To keep generality of our discussion, we introduce 
a parameter to model the velocity dispersion of DM per one direction, $\sigma_{\rm DM}$: 
\begin{align}
&\sigma_{{\rm PBH}y}^2=\sigma_{\rm DM}^2+\alpha^2(100)^2~({\rm km/s})^2\nonumber \\ 
&\sigma_{{\rm PBH}z}^2=\sigma_{\rm DM}^2+\alpha^2(100)^2~({\rm km/s})^2,
\end{align}
where we have again assumed the isotropic velocity dispersion. 
For our fiducial model, we assume $\sigma_{\rm DM}=220~{\rm km/s}$. We checked 
that a change in $\sigma_{\rm DM}$, say by $\pm 10\%$, gives only a small change in the following PBH constraints. 
Such a large velocity dispersion in the central region within the halo is supported by $N$-body simulation studies that simulate MW-scale halos, for example, 
Figure~2 of Ref.~\citep{2009MNRAS.395..797V}.

Table~\ref{tab:rho} gives a summary of the velocity distributions for stellar components in the bulge and disk regions and 
for PBHs (DM) in the halo region, respectively. 

\subsection{Mass spectrum of astrophysical lensing objects}
\label{sec:IMF}

\begin{figure}
\centering
\includegraphics[width=0.5\textwidth]{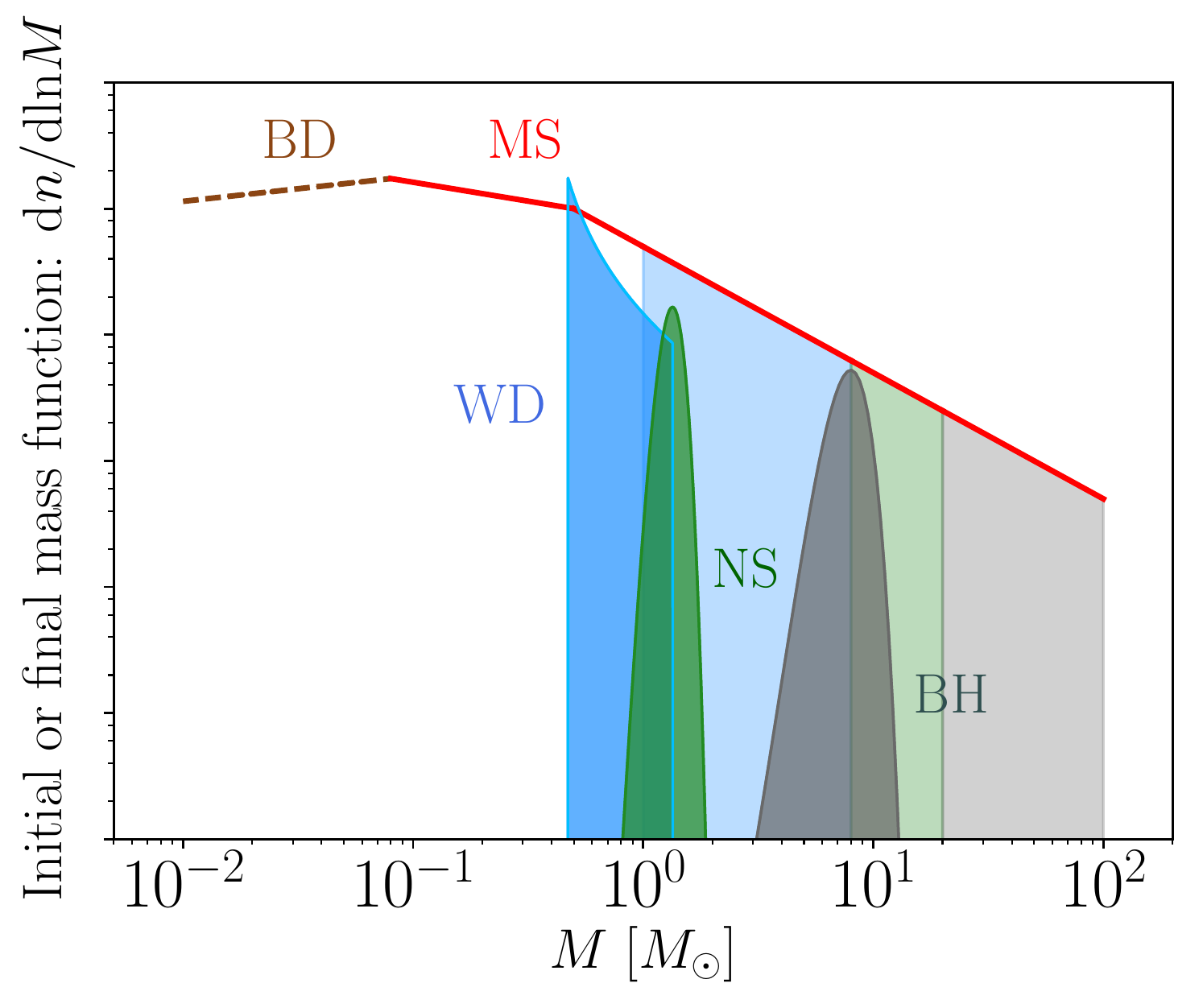}
\caption{Broken power-law curves denote the initial mass function of brown dwarfs (BD) and main-sequence star assuming 
the Kroupa-like model (see Eq.~\ref{eq:IMF}).
Note that $y$ axis is in an arbitrary scale. We assume that each massive star with $M\ge 1~M_\odot$ evolved into stellar remnant until 
today;
white dwarfs (WD) for stars with $1\le M/M_\odot\le 8$ following the initial and end mass relation, 
$M_{\rm WD}=0.339+0.129M_{\rm init}$, neutron stars (NS) for $8\le M/M_\odot \le 20$, and astrophysical black holes (BH)
for $M\ge 20~M_\odot$, respectively. For NSs and BHs, we assume a Gaussian for the end-mass function; we assume the Gaussian
with mean and 
width, $M_{\rm final}=1.33M_\odot$ and $\sigma=0.12M_\odot$ for NSs, while the Gaussian with $M_{\rm final}=7.8M_\odot$ and 
$\sigma=1.2M_\odot$ for BHs, respectively. The dark-shaded curves are the mass functions for WD, NS and BH, respectively. Because of the number conservation, the area under the curve for each stellar remnant, $\int\mathrm{d}\ln M~\mathrm{d}n/\mathrm{d}\ln M$, is the same as the area of the IMF over the corresponding range of initial main-sequence star masses (the two regions of similar color have 
the same area). For BDs, we determine the normalization of the mass function so that it matches the OGLE data at short timescales. 
}
\label{fig:IMF}
\end{figure} 
As implied by Eq.~(\ref{eq:t_E}), a timescale of microlensing light curve varies with a lens mass. To 
make a quantitative modeling of microlensing events as a function of the light curve timescale, we need to take into 
account the mass distribution of stellar lenses in the bulge and disk regions. 
Following \citet{2017Natur.548..183M} \citep[also see][]{2000ApJ...535..928G,Sumietal:11},
we consider brown dwarfs (BD), main sequence stars (MS), white dwarfs (WD), neutron stars (NS), and astrophysical black holes (BH)
as the constituents  in the bulge and disk regions.

To do this,  we first assume 
the Kroupa-like broken power-law {\it initial} mass function (IMF) for the stellar components \citep{2001MNRAS.322..231K}:
\begin{align}
&\frac{\mathrm{d} n_{\rm s}(M)}{\mathrm{d}\ln M}\nonumber\\
&=
\left\{
\begin{array}{ll}
A_{\rm BD}\left(\frac{M}{0.08M_\odot}\right)^{1-\alpha_{\rm BD}}& (0.01\le M/M_\odot \le 0.08)\\
A_{\rm MS}\left(\frac{M}{0.5~M_\odot}\right)^{1-\alpha_{\rm MS1}}& (0.08\le M/M_\odot \le 0.5)\\
A_{\rm MS}\left(\frac{M}{0.5~M_\odot}\right)^{1-\alpha_{\rm MS2}}& (M/M_\odot \ge 0.5)
\end{array} 
\right. ,
\label{eq:IMF}
\end{align}
for BD, low-mass MS stars ($0.08<M/M_\odot<0.5$), and high-mass MS stars ($M\ge 0.5M_\odot$), respectively; 
$A_{\rm BD}$ and $A_{\rm MS}$ are normalization parameters for which we will discuss below, and $\alpha_{\rm BD}, \alpha_{\rm MS1}$ and $\alpha_{\rm MS2}$ are the power-law index parameters for these components, respectively.
Following \citet{2017Natur.548..183M}
we assume $\alpha_{\rm BD}=0.8$, $\alpha_{\rm MS1}=1.3$ and $\alpha_{\rm MS2}=2$, respectively, where 
the slope for low-mass stars of $\simlt 1~M_\odot$ is taken from the study of the Galactic bulge IMF 
in Ref.~\citep{2000ApJ...530..418Z}. Throughout this paper we assume the same population composition 
of stellar components in the disk 
and bulge regions. 
The formation of BDs is still poorly understood \citep[see][for a review]{2001RvMP...73..719B}. Some fraction of BDs can be found in the planetary disk around a primary main-sequence star. Other population of BDs can form at the center of protoplanetary disk as a primary gravitating object of the system. Moreover, there might be some population of BDs ejected from the host system due to three-body scattering, which would be observed as ``free-floating planets'' in the interstellar space. A wide-orbit BD or a free-floating BD causes microlensing event characterized by the BD mass. If a lens system has both primary star and  BD in the close orbit, the microlensing event is characterized by the total mass (mainly the host star). Thus, as discussed in \citet{2017Natur.548..183M} \citep{Sumietal:11}, there is still a lot of discussion for the origin of microlensing events in a short timescale corresponding to BD masses or even shorter (smaller-mass) events. Hence, the amplitude of BD mass function is uncertain, and needs to be further study. 
In the following results, we will treat the BD normalization parameter $A_{\rm BD}$ as a free parameter, and determine it so that 
the model prediction matches the OGLE data in the corresponding short timescales.

Massive stars with masses $M_{\rm init}\ge 1~M_\odot$
have a rapid time evolution during the age of MW, and evolved into stellar remnants. 
Following \citet{2017Natur.548..183M}, we 
assume that all stars with initial masses $1\le M/M_\odot \le 8$ evolved into WDs following 
the empirical initial-final mass relation, $M_{\rm WD}=0.339+0.129M_{\rm init}$, after the mass loss; 
stars with $8\le M/M_\odot \le 20$ evolved into NSs for which we assume a Gaussian distribution with peak mass $M_{\rm final}=1.33M_\odot$ and width $\sigma=0.12M_\odot$ for the end masses; stars with $M\ge 20M_\odot$ evolved into astrophysical 
BHs for which we assume a Gaussian distribution with peak mass 
$M=7.8M_\odot$ and width $\sigma=1.2M_\odot$. 
We adopt the number conservation between initial stars and stellar remnants; each massive star evolved into each stellar remnant. 
Under this assumption, we found the ratio of the number of each stellar remnant relative to that of main sequence stars 
as
\begin{align}
{\rm MS}:{\rm WD}:{\rm NS}:{\rm BH}=1:0.15:0.013:0.0068~ .
\label{eq:ratio}
\end{align}
Throughout this paper we refer to stars with masses $0.08\le M/M_\odot\le 1$ as ``main sequence stars'' (MS). 

Fig.~\ref{fig:IMF} displays the model for the initial or final mass spectrum of BDs, MSs and stellar 
remnants, which we use in this paper. For the BD mass function, we will determine the normalization parameter so that the model prediction 
matches the OGLE microlensing events as we show below (we here adopt a normalization that is continuous with the stellar 
IMF at $M=0.08M_\odot$).
The number of each lensing population determines the frequency of microlensing. Then if we focus on the event rates for a particular 
light curve timescale, the events arise mainly from lensing objects of the corresponding mass scales (Eq.~\ref{eq:t_E}).
Thus, by studying the event rate as a function of 
the light curve timescales, one can distinguish contributions from different populations of lensing objects. 
Table~\ref{tab:massfunc} also gives the summary of our model for the mass spectrum of BDs, MS stars or 
stellar remnants.

Furthermore,  we assume the binary fraction $f_{\rm bin}=0.4$; the fraction of MS
stars or stellar remnants  are in binary systems. For simplicity we consider equal-mass binary systems: we
treat a microlensing of 
of a binary system by that of a lens with mass $M_{\rm binary}=2M$. 
We do not consider binary systems that contain two objects of different masses and contain two objects of different populations (e.g., 
MS-WD system) for simplicity. 
Consequently we decrease the number of lens systems from the above numbers in Fig.~\ref{fig:IMF} by the binary fraction. 
Including the binary systems give a slightly improved agreement between the model predictions and the OGLE data, 
but it is not an important assumption for our main results.

To perform a calculation of microlensing event rates, we need to specify the normalization parameter of MS IMF, $A_{\rm MS}$
 (Eq.~\ref{eq:IMF}). Recalling that the mass of Galactic bulge and disk regions is dominated by the total 
mass of low-mass MS stars, we determine $A_{\rm MS}$ by the condition 
\begin{align}
\rho_\ast=\int_{0.08M_\odot}^{M_\odot}\!\!\mathrm{d}\ln M~M\frac{\mathrm{d}n}{\mathrm{d}\ln M}.
\label{eq:norm}
\end{align}
Here $\rho_\ast$ is the normalization coefficient of mass density profile in the bulge and disk regions as given in Table~\ref{tab:model_summary}. With this normalization, $A_{\rm MS}$ has a dimension of $[{\rm pc}^{-3}]$. 
We assume the same composition of stars and stellar remnants everywhere in the disk and bulge regions; that is, we 
ignore a possible dependence of the composition on a position in the Galactic region. Details of our model are different from the model 
in \citet{2017Natur.548..183M}, 
so we will introduce a fudge normalization parameter later to model a possible uncertainty in the normalization: 
$A_{\rm MS}\rightarrow f_{\rm A}A_{\rm MS}$. However, we find $f_{\rm A}\simeq 1$, implying that our model is sufficiently close 
to the best-fit model in \citet{2017Natur.548..183M} or equivalently that the standard Galactic bulge/disk models are fairly accurate to
reproduce the observed timescale (mass) distribution of microlensing events as we will show below. 

\section{Results}
\label{sec:results}

\begin{figure*}
\centering
\includegraphics[width=0.8\textwidth]{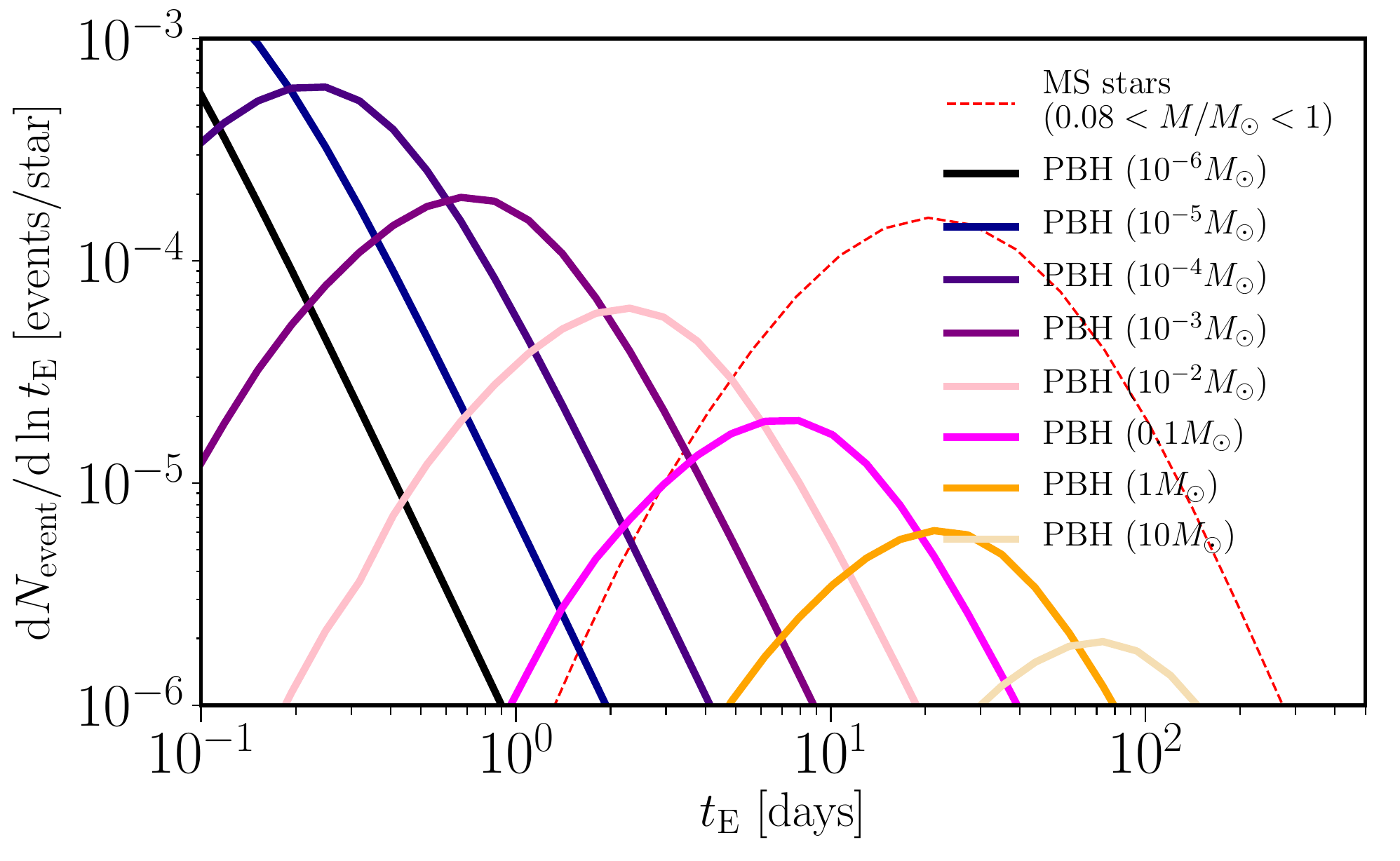}
\caption{The expected differential number of microlensing events per logarithmic interval of 
the light curve timescale $t_{\rm E}$, for a single star in the Galactic bulge region, assuming the 5-years observation as in the OGLE data. 
The quantity shown is defined in terms of the event rate described in Section~\ref{subsec:optical_depth} as 
$\mathrm{d}N_{\rm exp}/\mathrm{d}\ln t_{\rm E}\equiv 5~{\rm years}\times t_{\rm E}\times \mathrm{d}\Gamma/\mathrm{d}t_{\rm E}$
(see Eqs.~\ref{eq:gamma_b}, \ref{eq:gamma_d} and \ref{eq:gamma_pbh}).
Solid curves show the results for PBHs assuming that all DM in the MW region is made of PBHs of a given mass scale denoted 
by the legend: $f_{\rm PBH}=\Omega_{\rm PBH}/\Omega_{\rm DM}=1$. For comparison, dashed curve shows the result when main-sequence stars with mass in the range $[0.08,1]M_\odot$ are lenses, assuming the Galactic model for the star distribution in the bulge and disk regions. 
}
\label{fig:eventrate_pbh}
\end{figure*} 
\begin{figure}
\centering
\includegraphics[width=0.48\textwidth]{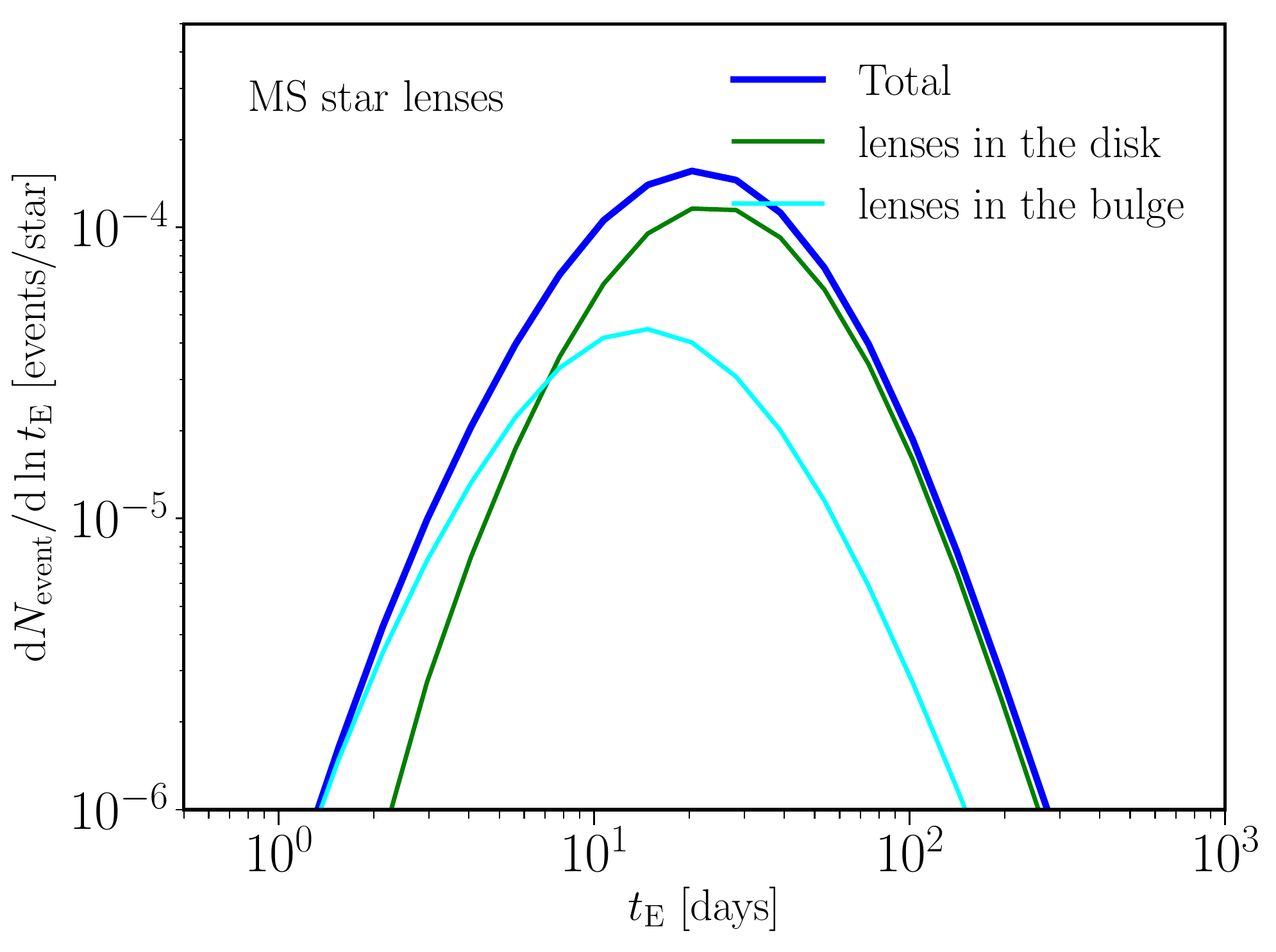}
\caption{Shown is the relative contributions of main-sequence stars in the bulge and disk regions to
the total event rates of microlensing in the previous figure.
}
\label{fig:eventrate_disk_vs_bulge}
\end{figure} 
\begin{figure}
\centering
\includegraphics[width=0.45\textwidth]{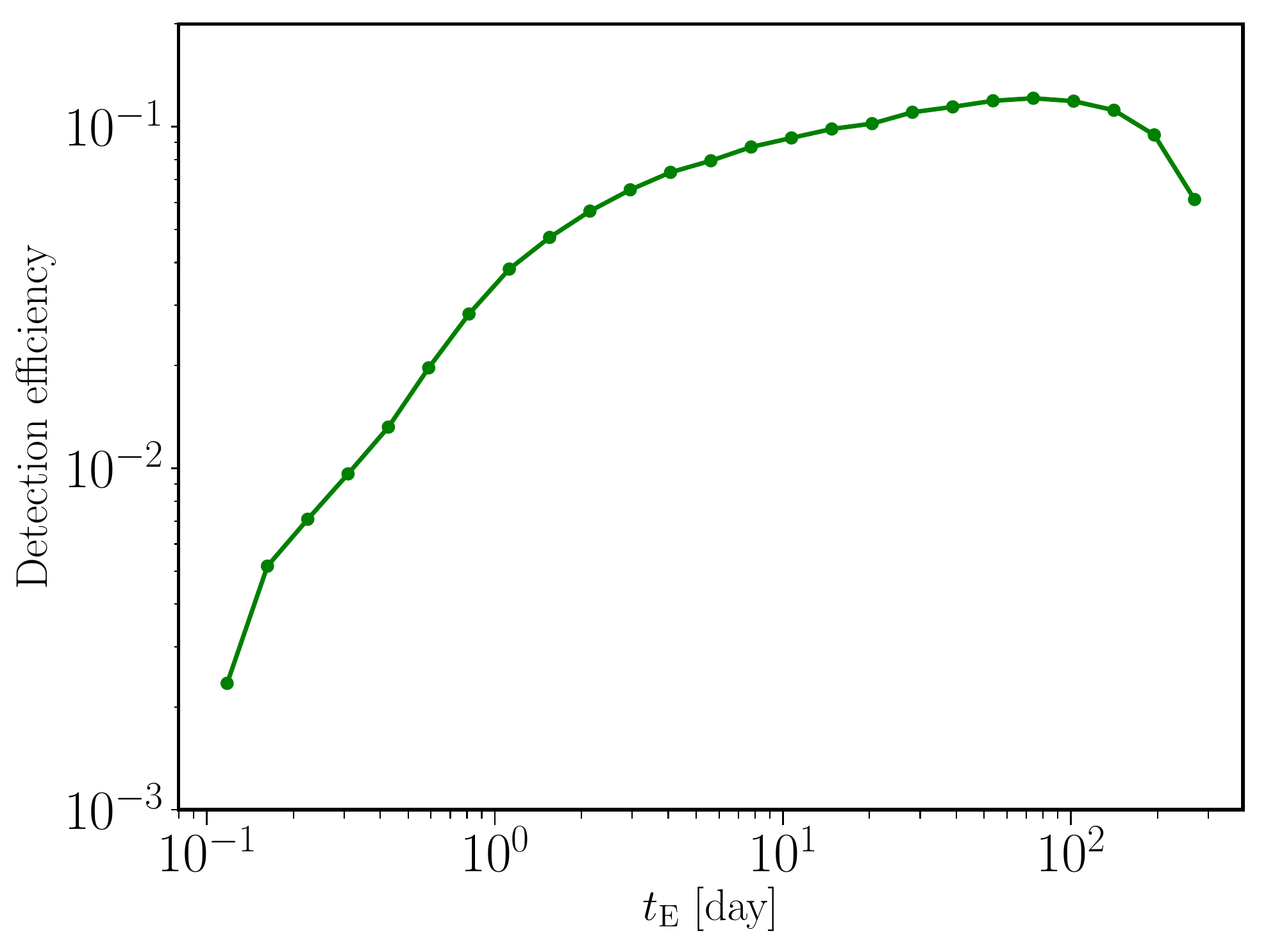}
\caption{The detection efficiency, $\epsilon(t_{\rm E})$, quantifying the probability that a mircolening event of timescale $t_{\rm E}$ is detected by the OGLE data. This represents a typical function that is taken from Extended Data Figure~2 in \citet{2017Natur.548..183M}.
}
\label{fig:efficiency}
\end{figure} 

\subsection{OGLE data}
\label{sec:ogle}

\begin{figure*}
\centering
\includegraphics[width=0.8\textwidth]{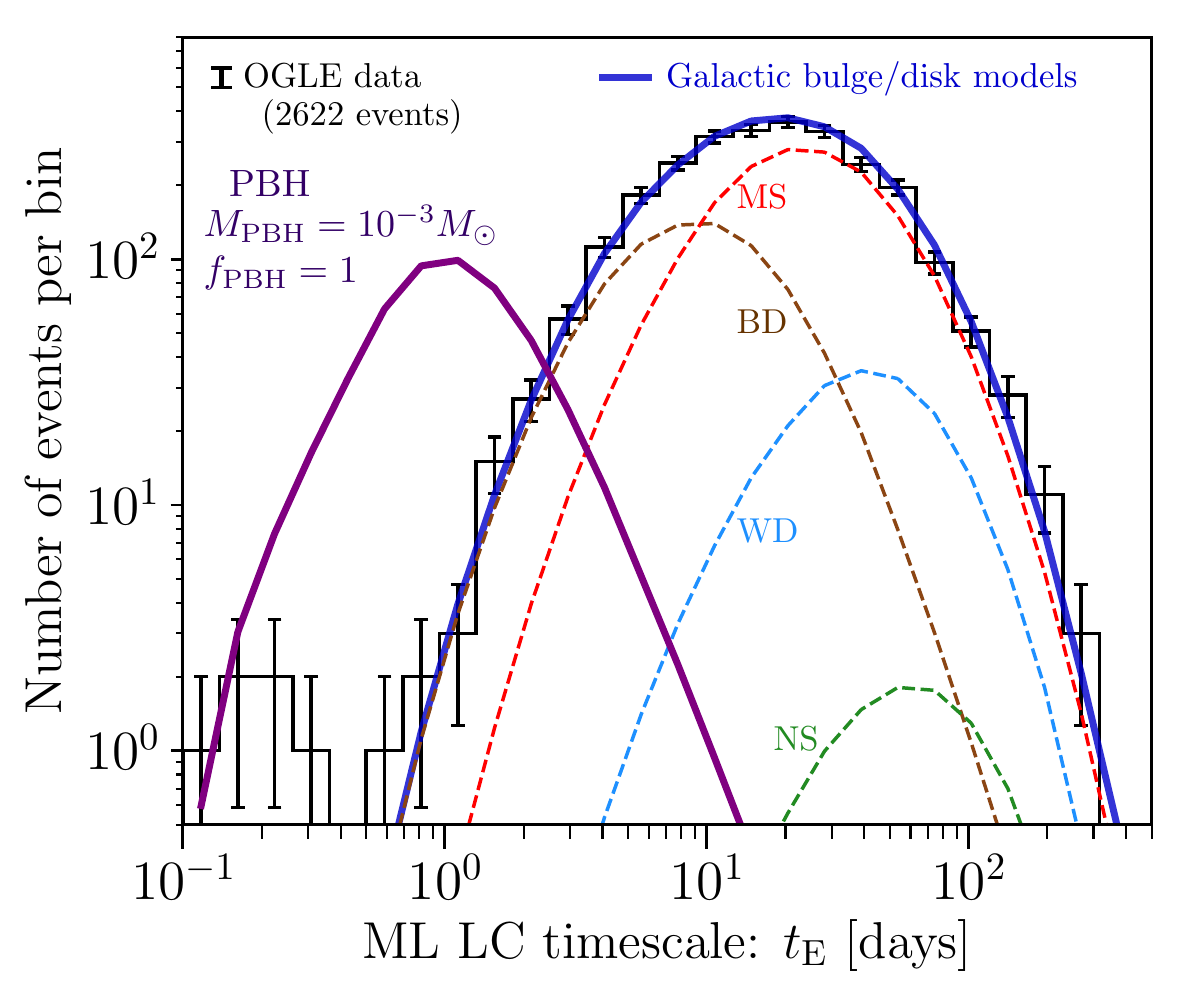}
\caption{Comparison of the 5-years OGLE data with the model predictions. The histogram
with errorbars denotes the OGLE data in each logarithmic bin of $t_{\rm E}$, where the errorbar is the $1\sigma$ Poisson uncertainties on the counts. The bold-blue solid curve shows the best-fit model assuming the stellar components in the bulge and disk regions. Other 
dashed curves show each contribution of brown dwarfs (BD), main sequence (MS) stars, white dwarfs (WD) and neutron stars (NS) to the total microlensing
events, respectively (see Fig.~\ref{fig:IMF}). The contribution of astrophysical black holes is outside the plotting range. 
As a demonstration, the purple curve shows the prediction if all DM is PBHs with mass $M_{\rm PBH}=10^{-3}M_\odot$ (Jupiter mass scales)
for $f_{\rm PBH}=1$. A sum of 
the PBH and astrophysical object contributions is too high compared to the OGLE events, and therefore such a PBH scenario
is ruled out by the OGLE data.
}
\label{fig:ogle_pdf}
\end{figure*} 
The results shown in this paper are all based on the microlensing data taken in the OGLE-IV sky survey during the 5 years, 2011--2015 \citep{2015AcA....65....1U,2017Natur.548..183M}. The OGLE survey uses the dedicated 1.3m Warsaw Telescope, located at Las Campanas Observatory, Chile. 
The OGLE survey carried out a long-term monitoring observation of the nine fields towards the Galactic bulge region
with a cadence of either 20~min or 60~min, covering 12.6 square degrees in total. 
After the careful cuts in a selection of microlensing events, the team created a catalog of 2622 microlensing events for each of which a timescale of the microlensing light curve ($t_{\rm E}$) is measured. Thus the OGLE datasets are quite rich and allow us to constrain the abundance and mass distribution of each lensing object population. As carefully studied in \citet{2017Natur.548..183M}, a majority of the OGLE events can be fairly well explained by superposition of microlensing events due to BDs, MS stars, and stellar remnants. Among these contributions, the origin of BDs is not well understood as we discussed \citep[see][for a review]{2001RvMP...73..719B}. 
Thanks to the unique power of microlensing that can probe a gravitational mass of a lensing object regardless of whether it is visible, the OGLE data in timescales less than $\sim 10~$days can be used to identify a microlensing contribution of BDs. However, the abundance 
and mass spectrum has not been fully understood yet. In addition to the BD contribution, those papers discussed a possible contribution 
of unbounded planets (wide-orbit planets or free-floating planets). \citet{2017Natur.548..183M} discussed that 
unbounded Jupiter-mass planets are  about 0.05 -- 0.25 planets per main sequence star, which is smaller
than previously advocated in Ref.~\citet{Sumietal:11}. The large OGLE dataset indicates even shorter timescale events that correspond
to unbounded Earth-mass planets. Interestingly, the OGLE data also indicates a "gap" (no microlensing event) at timescales between 
the BD or unbounded Jupiter-mass microlensing and the Earth-mass planets. Since planetary formation theory would predict a continuous mass spectrum, the gap, if real, seems very challenging to explain. For example, a mechanism preferentially scattering Earth-mass planets from the planetary system is needed. All these results are very interesting, and worth to further explore. 
All the datasets we use in this paper are taken from Extended Data Table in \citet{2017Natur.548..183M}.

In this paper, to derive PBH constraints, we employ the following two working hypotheses:
\begin{itemize}
\item[(1)] {\bf Null hypothesis of PBH microlensing}:  we assume that all the OGLE microlensing events are due to astrophysical objects, i.e.
 BDs, stars and remnants, so do not contain any PBH microlensing event. Then we use all the OGLE events to obtain an upper limit on the abundance of PBHs assuming the monochromatic mass spectrum. 
\item[(2)] {\bf PBH hypothesis of the 6 ultrashort-timescale OGLE events, $t_{\rm E}\simeq[0.1,0.3]~{\rm days}$}. The OGLE data found the 6 ultrashort-timescale mcirolensing events (Fig.~\ref{fig:ogle_pdf}), which seem a different population from the majority of events in the longer timescales due to BDs, stars and remnants. The timescale implies an Earth-mass lens. 
Although the origin might be unbounded Earth-mass planets, 
we here assume that 
the ultrashort-timescale events are due to PBHs, but other longer timescale events are due to astrophysical objects as in 
case (1).  Under this hypothesis, we
derive an allowed region of PBHs in two parameter space of the abundance and mass scale, assuming the monochromatic mass spectrum. 
\end{itemize}

\subsection{Event rate of microlensing}
We are now in a position to compute event rates of microlensing by plugging the model ingredients, which 
we have discussed up to the preceding subsection, into the equations such as Eq.~(\ref{eq:gamma_b}).

Fig.~\ref{fig:eventrate_pbh} shows the expected differential number of microlensing events per logarithmic interval 
of the light curve timescale $t_{\rm E}$, for a single source star in the bulge region, assuming the 5-years observation as in the 
OGLE data. For PBH microlensing, we adopt the model ingredients in Sections~\ref{sec:density_velocity} for the mass density profile 
and velocity distribution, assuming the monochromatic mass scale. We assumed that all DM
is made of PBHs of each mass scale: $f_{\rm PBH}=1$. If we consider lighter-mass PBHs, the number density of PBHs increases and such PBHs yield a higher frequency of 
microlensing events with shorter timescales. 
In particular, for microlensing events with timescales shorter than a few days, PBHs with $M_{\rm PBH}\simlt 10^{-1}M_\odot$ could produce a larger number of microlensing events than MS stars of $\sim 1~M_\odot$ do, if such PBHs constitute a significant fraction of DM. 

In Fig.~\ref{fig:eventrate_disk_vs_bulge} we study relative contributions of MS stars in the bulge and disk regions to the total of MS microlensing events. It can be found that stars in the disk region gives a dominant contribution, while the bulge star contribution is significant for shorter timescale events. 

\subsection{Comparison with the 5-years OGLE data}

We now compare the model predictions of microlensing events with the 5-years OGLE data. The OGLE data contains 2622 events over the range of light curve timescales, $t_{\rm E}=[10^{-1}, 300]~{\rm days}$ (see Extended 
Data Table~4 in \citet{2017Natur.548..183M}). The expected number of microlensing events per a given timescale interval of $[t_{\rm E}-\Delta t_{\rm E}/2, t_{\rm E}+\Delta t_{\rm E}/2]$ is computed as
\begin{align}
N_{\rm exp}(t_{\rm E})=t_{\rm obs}N_{\rm s}f_{\rm A}\int_{\rm t_{\rm E}-\Delta t_{\rm E}/2}^{t_{\rm E}+\Delta t_{\rm E}/2}
\!\!\mathrm{d}\ln t_{\rm E}^\prime \frac{\mathrm{d}^2\Gamma}{\mathrm{d}\ln t_{\rm E}^\prime
}\epsilon(t_{\rm E}^\prime),
\label{eq:N_exp}
\end{align}
where $t_{\rm obs}$ is the total observation time, $N_{\rm s}$ is the total number of source stars in the OGLE bulge fields, and 
$\epsilon(t_{\rm E})$ is the ``detection efficiency'' quantifying the probability that a microlensing event of timescale $t_{\rm E}$ is detected 
by the OGLE data. 
For the OGLE data, $t_{\rm obs}=5~{\rm years}$ and $N_{\rm s}=4.88\times 10^7$ \citep[see Extended Data Table~2 in][]{2017Natur.548..183M}. We employ the detection 
efficiency, $\epsilon(t_{\rm E})$, that is taken from Extended Data Figure~2 in \citet{2017Natur.548..183M}, which is explicitly shown in Fig.~\ref{fig:efficiency}.
We do not include variations of the detection efficiency in the different OGLE fields for simplicity.
The coefficient $f_{\rm A}$ is a fudge normalization factor that takes into account a possible difference in details of our model calculations and the model of \citet{2017Natur.548..183M}.

In Fig.~\ref{fig:ogle_pdf} we compare the model prediction of microlensing event rates with the 5-years OGLE data. First of all, 
a majority of the OGLE microlensing events has a single peak around the timescale, $t_{\rm E}\sim 20~{\rm days}$, and has a gradual decrease at the shorter and longer timescales than the peak timescale. Thus the OGLE data suggests only a single population of the underlying lensing objects, except for the 6 ultrashort-timescale events, $t_{\rm E}=[0.1,0.3]~{\rm days}$, which we will discuss later. Interestingly, the model assuming the standard Galactic bulge and disk models (see Section~\ref{sec:density_velocity}) 
can fairly well reproduce event rates for the main population of OGLE microlensing events. Furthermore, by employing the mass distribution of BD, stars and stellar remnants, the model can reproduce the distribution of light curve timescales 
(see Section~\ref{sec:IMF}). Although we introduced a fudge factor 
in Eq.~(\ref{eq:N_exp}) to model a possible difference between our model and the model in \citet{2017Natur.548..183M}, we found $f_{A}=0.99$ to have a nice agreement of our model prediction with the OGLE data at timescales greater than the peak timescale. 
The best-fit $f_{\rm A}$ value is close to unity, and reflects the fact that the Galactic bulge 
and disk model, constructed based on observations and the previous knowledges, is fairly accurate. 
As we discussed, the origin and nature of unbounded BDs, which cause shorter timescale events, is poorly understood. 
We found that the model matches the microlensing events at timescales shorter than the peak timescale, if we assume 0.18 BDs per main-sequence star (see Table~\ref{tab:massfunc}). 
The figure clearly shows that MS stars with $0.08\le M/M_\odot\le 1$ give a dominant contribution to the OGLE events at timescales, $t_{\rm E}\simgt 10~{\rm days}$,
while stellar remnants give secondary contributions. 
BDs give a dominant contribution at the shorter timescales. However, the figure shows that, as long as we assume a smooth model for the density and velocity distributions of BDs, the model cannot reproduce the ultrashort-timescale events of $t_{\rm E}\sim 0.1~{\rm days}$. This clearly indicates a distinct, second population of small-mass lensing objects. 
On the other hand, PBHs do not necessarily follow the similar timescale distribution of microlensing events to that of BD, stars 
or remnants, because DM has different spatial and velocity distributions from the stellar populations. As an example, Fig.~\ref{fig:ogle_pdf} shows the result for case that PBHs with mass $M_{\rm PBH}=10^{-3}M_\odot$ are DM. A sum of the PBH and stellar population contributions give too many microlensing events compared to the OGLE data. In other words, such a PBH population is not allowed by the OGLE data. Thus we can use the OGLE data to obtain an upper bound on the abundance of PBHs with varying PBH mass scales. 

\subsection{Upper bound on the PBH abundance under null hypothesis}
\label{sec:upper_bound}
\begin{figure}
\centering
\includegraphics[width=0.48\textwidth]{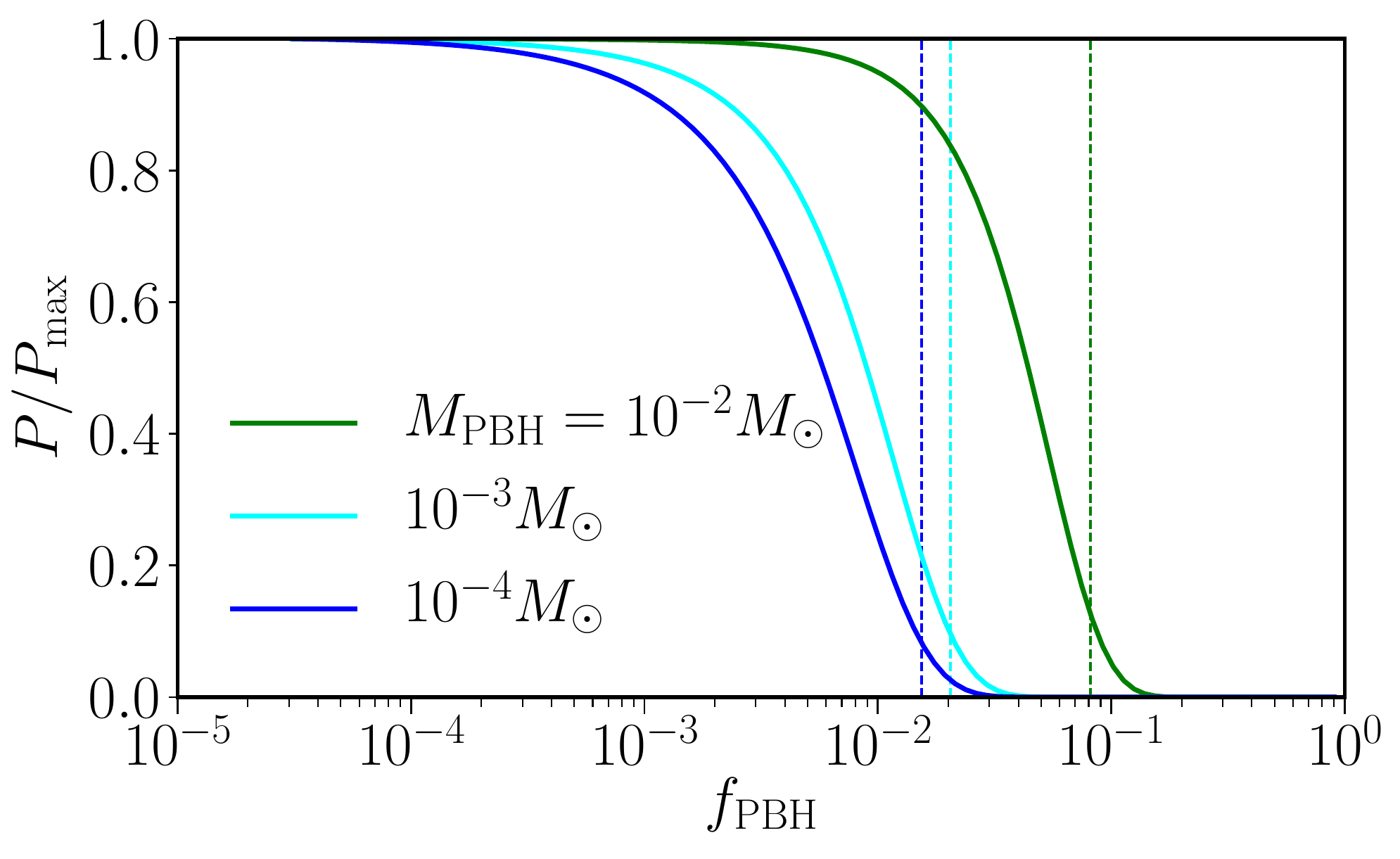}
\caption{Posterior distribution of the $f_{\rm PBH}$ parameter (the PBH mass fraction to DM) assuming ``null hypothesis'' that there is no PBH microlensing in the OGLE data (see text for details). Here we 
show, as examples, three cases for PBH mass scale; $M_{\rm PBH}/M_\odot=10^{-4}, 10^{-3}$ or $10^{-2}$, respectively, 
which is computed from Eq.~(\ref{eq:posterior}) by comparing the model prediction of PBH microlensing event rates with the OGLE data. 
The vertical dashed line for each curve denotes 
 95\% CL upper limit on the abundance of PBH for each mass case, which is obtained by computing the integration of the posterior distribution, 
 $\int^{f_{\rm PBH, 95\%}}_0\!\mathrm{d}f_{\rm PBH}~P(f_{\rm PBH})=0.95$.
	\label{fig:posterior}
}
\end{figure}
\begin{figure*}
\centering
\includegraphics[width=0.9\textwidth]{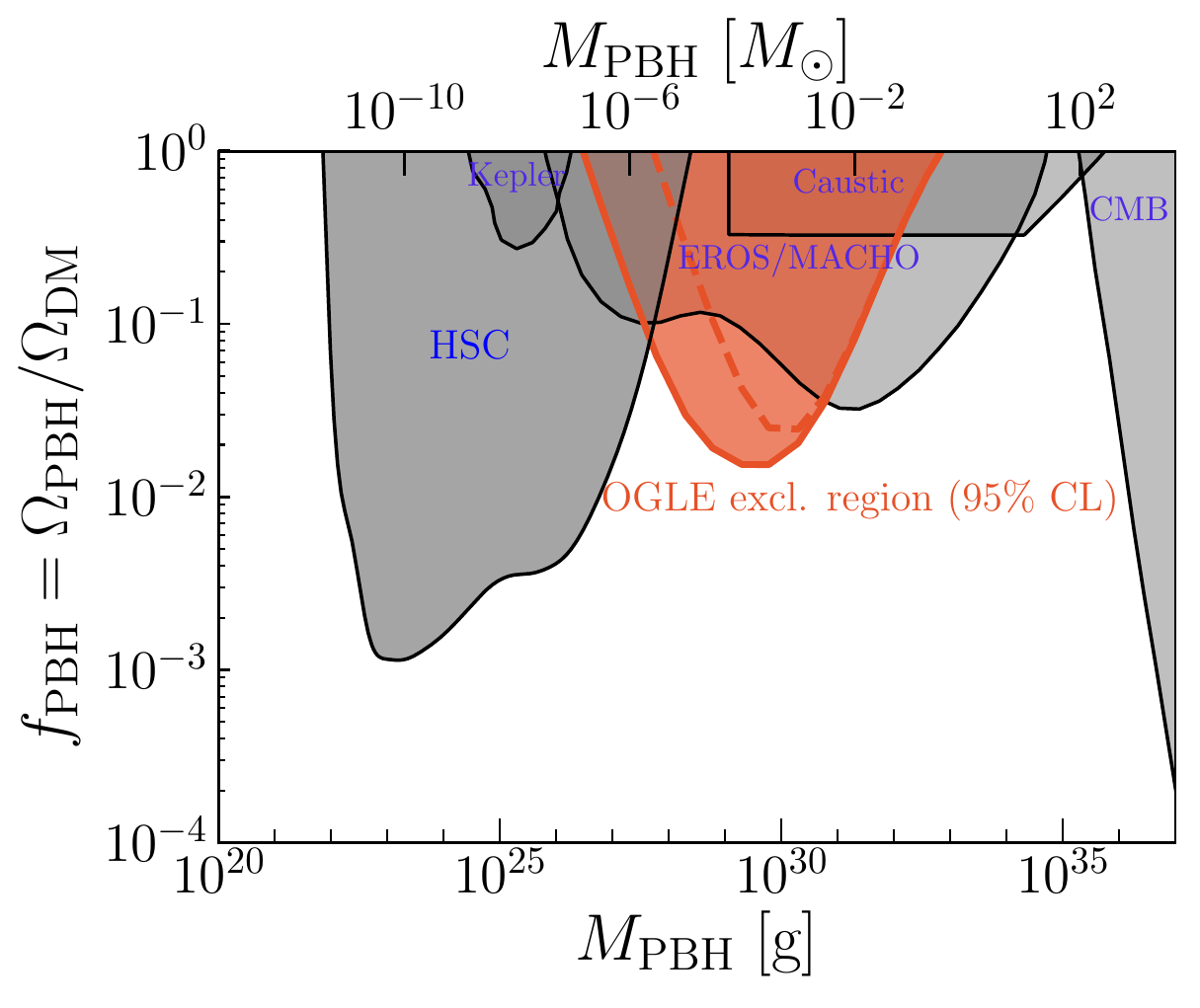}
\caption{Red shaded region corresponds to the 95\% C.L. upper bound on the PBH mass fraction to DM, derived assuming the null hypothesis that there is no PBH microlensing event in the 5-year OGLE data (see text for details).
Here we assume a monochromatic mass function of PBHs, and we derive the upper bound at each mass scale denoted in the $x$-axis. 
The dashed curve shows the upper bound if the OGLE data in the 4 shortest timescale bins for the ultrashort-timescale events 
is not used for the null hypothesis. 
This constraint can be compared with other observational constraints as shown by the gray shaded regions: 
the microlensing search of stars in the Andromeda galaxy from the one-night Subaru Hyper Suprime-Cam data (``HSC'') \citep{2017arXiv170102151N}, the mirolensing search from the 2-years Kelper data (``Kepler'') \citep{Griestetal:14},
the earlier MACHO/EROS/OGLE microlensing search (``EROS/MACHO'') \citep{EROS:07}, the microlensing of extremely magnified stars near caustics of a galaxy cluster (``Caustics'') \citep{2018PhRvD..97b3518O}
and the accretion effects on the CMB observables (``CMB'') \citep{Ali-HaimoudKamionkowski:17}, which is the result updated from the earlier estimate \citep{Ricottietal:08}. 
}
\label{fig:upper_bound}
\end{figure*} 
\begin{figure*}
\centering
\includegraphics[width=0.9\textwidth]{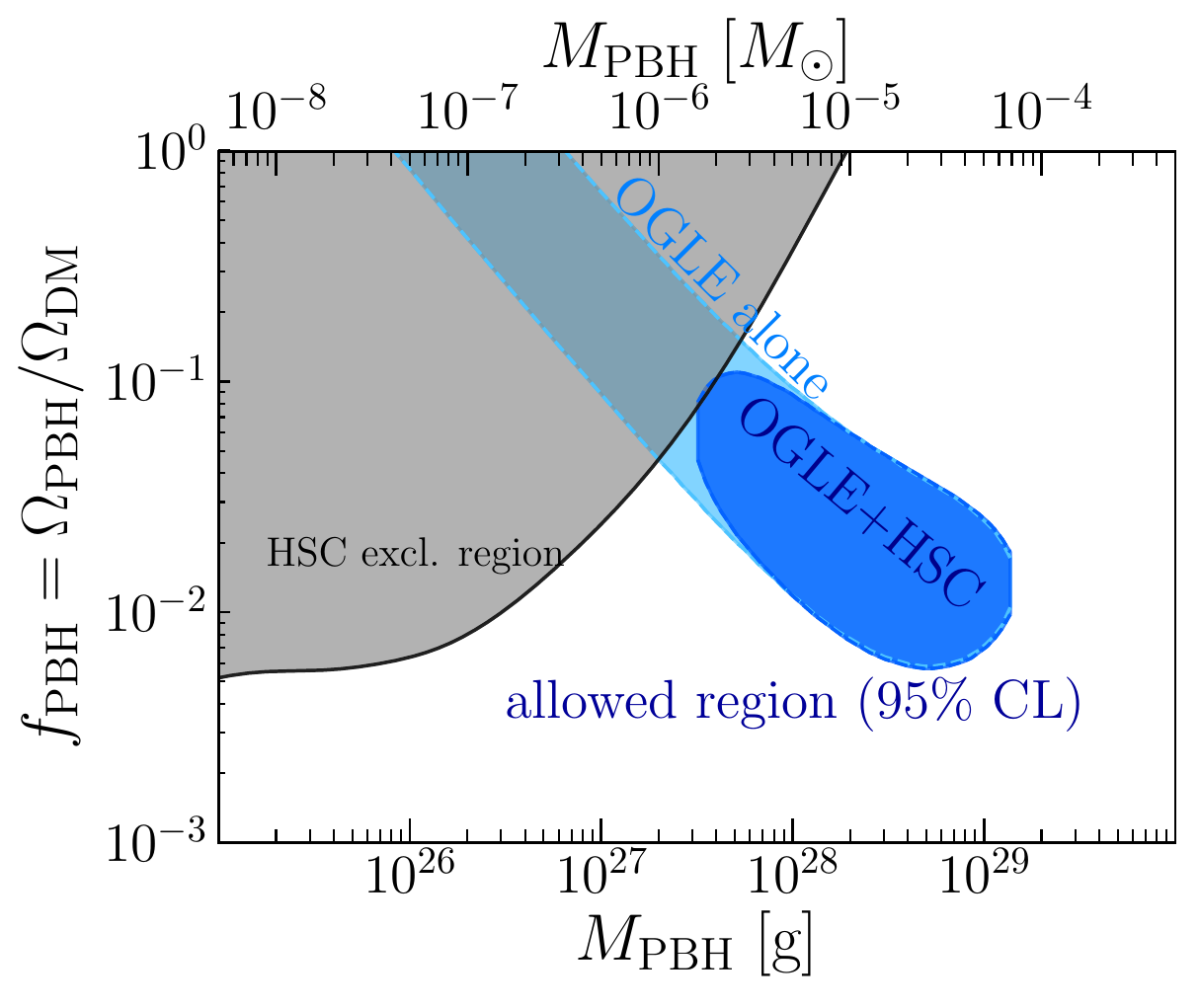}
\caption{Shaded blue region is the 95\% CL allowed region of PBH abundance, obtained by assuming that 6 ultrashort-timescale microlensing 
events in the OGLE data are due to PBHs. Note that we assume a monochromatic mass scale for PBHs as given in the $x$-axis. The allowed region is computed from the condition $P(f_{\rm PBH}, M_{\rm PBH})/P_{\rm max}>0.046$, which corresponds to 95\% CL if the surface of posterior distribution follows a two-dimensional Gaussian distribution ($P_{\rm max}$ is the posterior distribution for the best-fit model). Dark shaded region shows the result when combining the allowed region of the ultrashort-timescale events with the upper bounds from the Subaru constraints and the longer timescale OGLE data. 
}
\label{fig:allowed_region}
\end{figure*} 
As we showed in the preceding subsection, the Galactic bulge and disk models including the stellar components fairly well
reproduce the OGLE data except for the 6 ultrashort-timescale events. In other words, the OGLE data does not necessarily imply the existence of PBH microlensing in the data. 
As the first working hypothesis (see Section~\ref{sec:ogle}), 
we here employ ``null hypothesis'' to obtain an upper limit on the abundance of PBHs. That is, we assume that all the observed OGLE microlensing events, including 
the 6 ultrashort-timescale events, are due to the stellar components, or equivalently there is no PBH lensing in the OGLE data. 
This would give us a most stringent upper bound on the PBH abundance. If we allow a possible PBH contribution to the OGLE data in addition to the stellar events, it would give us a more relaxed upper bound or could even allow for a detection of PBH.
However, this requires a perfect knowledge of the Galactic stellar components, which is not straightforward. 

\begin{figure}
\centering
\includegraphics[width=0.47\textwidth]{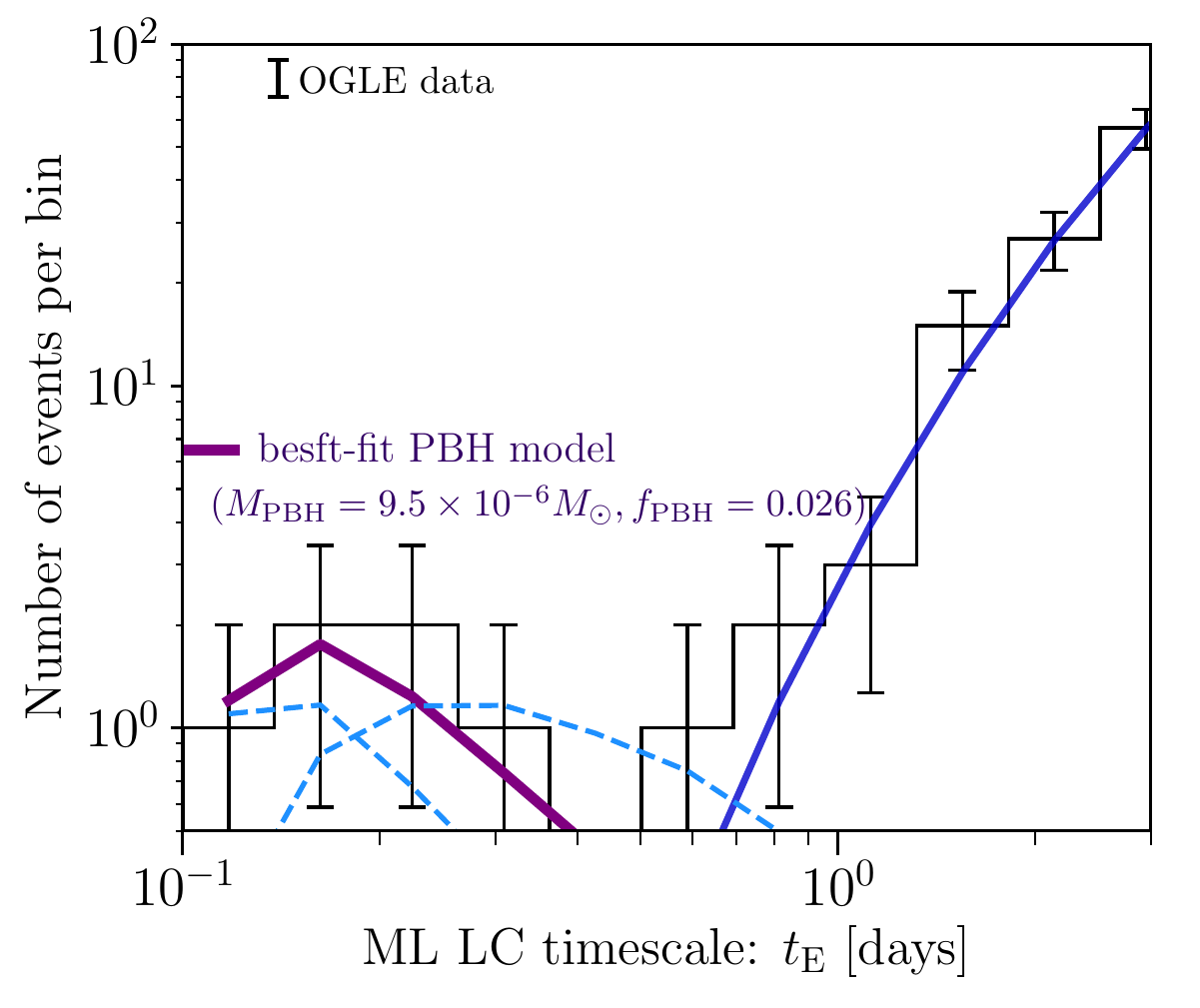}
\caption{Bold purple-solid line shows the best-fit PBH model in each of the 4 shortest timescale bins, which shows a good agreement with the distribution of 6 ultrashort-timescale OGLE events. The best-fit model is characterized by 
$M_{\rm PBH}=9.5\times 10^{-6}M_\odot$ and $f_{\rm PBH}=0.026$. For comparison, the two dashed lines show the predictions for two models that are close to the boundary of the allowed region of ``OGLE+HSC'' in Fig.~\ref{fig:allowed_region}; 
 $(M_{\rm PBH},f_{\rm PBH})=(1.6\times 10^{-6},0.062)$ or ($6.9\times 10^{-5},0.014$), respectively. These models do not give a good match to the timescale distribution of the ultrashort-timescale events, and also become inconsistent with the upper bounds of the HSC M31 and/or the longer timescale OGLE data. 
}
\label{fig:bestfit_allowed}
\end{figure} 
\begin{figure}
\centering
\includegraphics[width=0.47\textwidth]{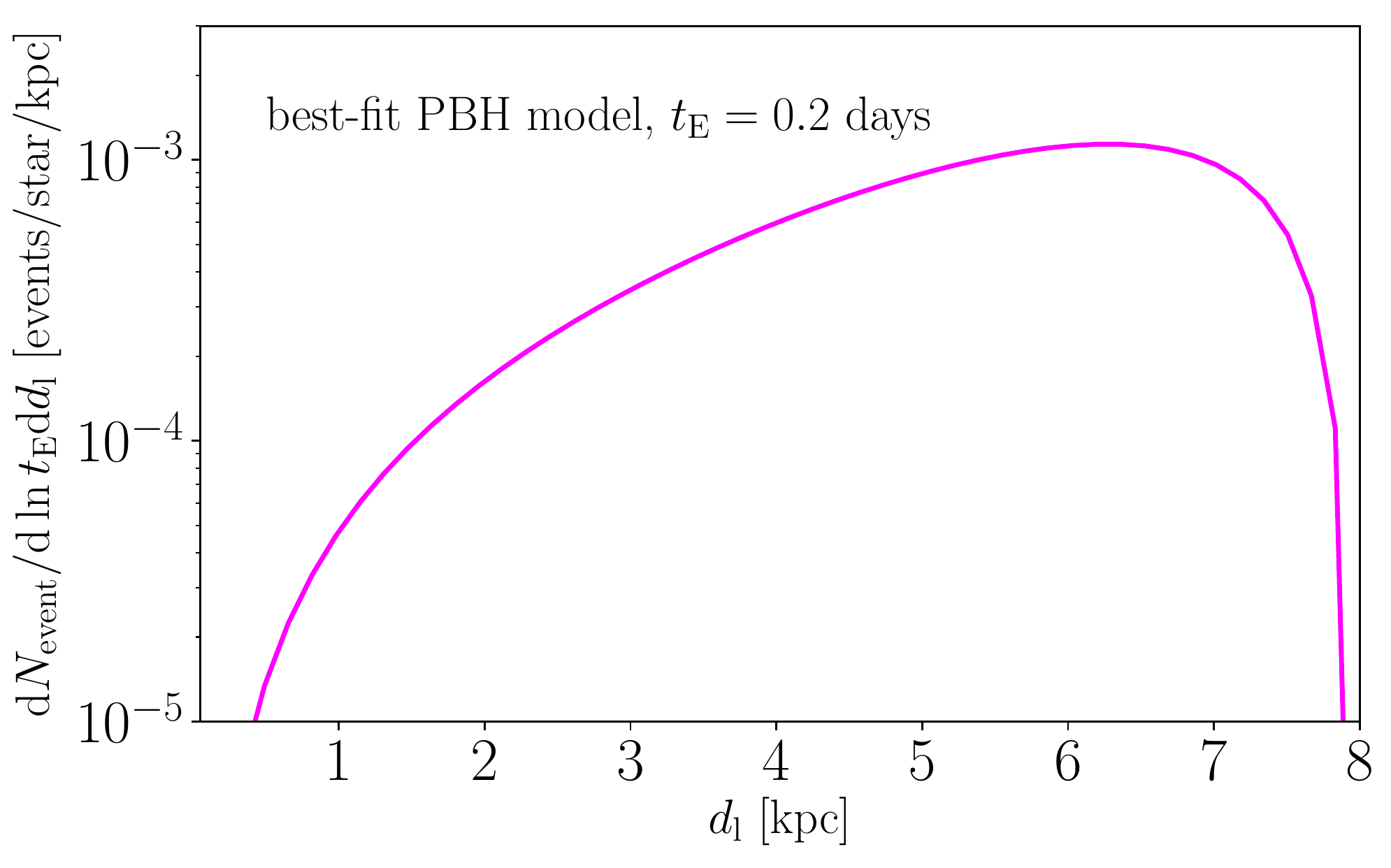}
\caption{The contribution of PBHs at each distance to the total event rate of timescale $t_{\rm E}=0.2~{\rm days}$. 
Here we consider the best-fit PBH model ($M_{\rm PBH}=9.5\times 10^{-6}M_\odot$, $f_{\rm PBH}=0.026$) to the ultrashort-timescale OGLE 
events in the previous figure. } 
\label{fig:dN_dl}
\end{figure} 
We assume that the OGLE counts of microlensing events at each timescale bin follows the Poisson distribution. This is a good assumption because the same lensing object very unlikely produces multiple lensing events (lensing for multiple source stars)
because of smallness of the lensing optical depth, $\tau\sim 10^{-6}$
 (see Table~\ref{tab:model_summary}). Hence we can safely assume that different microlensing events are independent and uncorrelated with each other. Under these assumptions, we assume that the log likelihood of 
OGLE microlensing events is given by 
\begin{align}
\ln{\cal L}({\bf d}|{\bm \theta})=
\sum_{i=1}^{n_{\rm bin}}\left[
N_{\rm obs}(t_{{\rm E},i})\ln \lambda(t_{{\rm E},i})-\lambda(t_{{\rm E},i})-\ln N_{\rm obs}(t_{{\rm E},i})!
\right]
\label{eq:likelihood}
\end{align}
where 
$N_{\rm obs}(t_{{\rm E},i})$ is the observed number of events at the $i$-th timescale bin ($t_{{\rm E},i}$); 
{\bf d} is the data vector, ${\bf d}\equiv \{N_{\rm obs}(t_{{\rm E},1}), N_{\rm obs}(t_{{\rm E},2}), \dots, 
N_{\rm obs}(t_{{\rm E},n_{\rm bin}})\}$ in our case, $n_{\rm bin}$ is the number of timescale bins ($n_{\rm bin}=25$ as can be found from  Fig.~\ref{fig:ogle_pdf}); ${\bm \theta}$ is the model vector; 
$\lambda(t_{{\rm E},i})$ is the expectation number of events at the bin. When we include PBH microlensing contributions, we model the expectation number by 
\begin{align}
\lambda(t_{{\rm E},i})=N_{\rm obs}(t_{{\rm E},i})+N^{\rm PBH}_{\rm exp}(t_{{\rm E},i}) .
\label{eq:poisson_lambda}
\end{align}
Here $N_{\rm exp}^{\rm PBH}(t_{{\rm E},i})$ is the expected number of microlensing events due to PBHs at the $i$-th 
timescale bin, which is computed from Eq.~(\ref{eq:N_exp}) once the PBH mass fraction to DM, $f_{\rm PBH}$, is specified 
for an assumed PBH mass scale ($M_{\rm PBH})$; $N_{\rm exp}^{\rm PBH}({{\rm E},i})\propto f_{\rm PBH}$. 
As a conservative approach, we use the observed counts, $N_{\rm obs}(t_{{\rm E},i})$ for the expectation value of microlensing events
due to stellar components. 
In the following, we assume that the MW DM model for the spatial and velocity distributions for PBH in Sections~\ref{sec:density_velocity}, 
and we treat 
the PBH mass fraction parameter, $f_{\rm PBH}$, as a free parameter for an assumed PBH mass scale ($M_{\rm PBH}$). Namely we 
consider a single model parameter for an assumed PBH mass scale (we will discuss later for a possible extension of this assumption). 
When $f_{\rm PBH}=0$, i.e. $N_{\rm exp}^{\rm PBH}=0$, the maximum likelihood is realized because of $N^{\rm PBH}_{\rm exp}\ge 0$.
The last term in the above log likelihood is irrelevant for parameter inference, because it is a fixed number irrespectively of model parameter $(f_{\rm PBH})$.

Given the likelihood function and the PBH model (denoted as ${\cal M}$), the posterior distribution of model parameter, $f_{\rm PBH}$, is computed based 
on the Bayes's theorem as
\begin{align}
P(f_{\rm PBH}|{\bf d},{\cal M})=\frac{{\cal L}({\bf d}|f_{\rm PBH})\Pi(f_{\rm PBH})}{P({\bf d}|{\cal M})},
\label{eq:posterior}
\end{align}
where $\Pi(f_{\rm PBH})$ is a prior of $f_{\rm PBH}$ and $P({\bf d}|{\cal M})\equiv {\cal E}$ is the evidence. In this paper, 
we assume a flat prior, $f_{\rm PBH}\le 1$; the total PBH mass in the MW region cannot exceed the DM mass. 
By computing the above equation with varying the model parameter $f_{\rm PBH}$, we can obtain the posterior distribution for an
assumed mass scale of PBH. Fig.~\ref{fig:posterior} shows some examples for the posterior distribution for a given PBH mass scale, obtained from the above method. 

Fig.~\ref{fig:upper_bound} shows 95\% CL upper bound on the PBH abundance at different mass scales. The OGLE data improves the constraints on the abundance for PBHs in the mass range $M_{\rm PBH}\simeq [10^{-3},10^{-6}]$.
The dashed curve shows the upper bound if we do not include the 6 ultrashort-timescale events in the 4 shortest timescale bins for the null hypothesis. Thus the upper bound at small mass scales is sensitive to the assumption of whether we include the short timescale bins in the analysis. 
The results 
can be compared with other constraints such as those from the Subaru Hyper Suprime-Cam (HSC \footnote{\url{https://hsc.mtk.nao.ac.jp/ssp/}}) observation of Andromeda galaxy (M31) \citep{2017arXiv170102151N} and the earlier MACHO/EROS experiments \citep{Alcocketal:00,EROS:07}. The OGLE bound is stronger than that of MACHO/EROS due to the larger sample 
of microlensing events. The OGLE constraint is complementary to the Subaru HSC result that uses even denser (2~min) cadence data of M31 to search for microlensing for a larger number of source stars, but from only a single night observation. If we want to extend the constraint to PBHs at larger mass scales, we need to use the microlensing data extending to longer timescales such as year timescales. 

\subsection{A possible detection of Earth-mass scale PBHs from short-timescale OGLE data}

Now we employ the second working hypothesis in Section~\ref{sec:ogle}. That is we consider a case that the 6 ultrashort-timescale OGLE events, 
in $t_{\rm E}\simeq [0.1,0.3]~{\rm days}$, are due to PBH microlensing.  In this case, we assume that the 
expectation number of microlensing events at each of the first 4 timescale bins (in the ultrashort-timescale bins) is given as
\begin{align}
\lambda(t_{{\rm E},i})=N^{\rm PBH}_{\rm exp}(t_{{\rm E},i})
\end{align}
for the Poisson distribution of micrlensing event counts (Eq.~\ref{eq:likelihood}). Here we should again note that we assume the monochromatic mass scale for PBHs. We also assume that the OGLE events in the longer timescales, i.e. the majority of OGLE 
events, are not due to PBHs (i.e no PBH microlensing as in the preceding section).

Fig.~\ref{fig:allowed_region} shows a 95\% CL {\it allowed} region of PBHs in two parameter space of its mass and abundance. The thin-blue shaded region corresponds to the allowed region obtained from the OGLE data alone, while the thick shaded region is the 
allowed region when combining the OGLE results with the null hypothesis of PBH lensing for the longer timescale OGLE events ($t_{\rm E}\simgt 0.5~{\rm days}$) and the the HSC constraints for M31.
The best-fit model, which has a maximum likelihood, is a model with $M_{\rm PBH}=9.5\times 10^{-6}M_\odot$ and $f_{\rm PBH}=0.026$.  
The figure shows that PBHs of Earth-mass scales 
($3\times 10^{-6}M_\odot$) can well reproduce the 6 ultrashort-timescale OGLE events if the abundance is at a per cent level. 
This allowed region is also consistent with null PBH results in the HSC data and the longer timescale OGLE data (if assuming the monochromatic mass spectrum).

Fig.~\ref{fig:bestfit_allowed} compares the best-fit PBH model with the timescale distribution of OGLE events in the shortest timescales. 
%
%
Interestingly, 
the width 
of OGLE timescale distribution is nicely reproduced by the velocity distribution of PBHs in the MW DM model.   
For comparison, the two dashed curves denote the model predictions for two models that are close to the boundary of 95\% C.L. intervals in the allowed region, which are specified by model parameters 
$(M_{\rm PBH}/M_\odot,f_{\rm PBH})=(1.6\times 10^{-6},0.062)$ or $(6.9\times 10^{-5},0.014)$, respectively. These failed models under- or over-predict the microlensing event rats over the range of lightcurve timescale bins, and also become inconsistent
with the upper bounds of the HSC M31 data and the longer timescale OGLE data. This result implies that, if PBHs have a wide mass spectrum extending to larger masses than the best-fit mass, such a model generally fills the gap around $t_{\rm E}\simeq 0.4~{\rm days}$ between the ultrashort-timescale events and the main population.  
Note that PBH models with smaller masses than the best-fit mass generally predict too many microlensing events at even shorter timescales $t_{\rm E}\le 0.1~{\rm days}$, however, the OGLE data does not have a sensitivity due to the limitation of the cadence data (20~min cadence), and therefore the OGLE data has no sensitivity to events at 
$t_{\rm E}\le 0.1~{\rm days}$, which is taken into account by the detection efficiency in Fig.~\ref{fig:efficiency}. The original event rates can have an increasing function at $t_{\rm E}\simlt 0.1~{\rm days}$. These mass-scale PBHs are well constrained by the HSC results. 

As we have shown, the PBH model assuming the MW DM model can give an alternative explanation of the ultrashort-timescale OGLE events. Since there is an uncertainty in the MW DM model, especially the DM distribution around the halo center, one might worry whether such an uncertainty in the PBH distribution around the halo center is sensitive to our results. Fig.~\ref{fig:dN_dl} shows which distant PBHs for the best-fit model contribute the microlensing events at timescale $t_{\rm E}=0.2~{\rm days}$. Due to lensing efficiency, PBHs over a wide range of distances between the Earth and the Galactic center equally contribute the lensing events. The PBHs near the Galactic center, where the DM distribution is most uncertain, is not particularly sensitive to the final result. This lens distance dependence is contrary to DM annihilation that is quite sensitive to details of the DM density in the Galactic center.

\section{Discussion and Conclusion}
\label{sec:discussion}

In this paper we have used the largest sample of microlensing events for stars in the Galactic bulge, obtained from the 5-years OGLE observation, to constrain the abundance of PBH that is a viable candidate of DM in the MW region. The 2622 microlensing data contains rich information on event rates and light curve timescales that correspond to the abundance and mass scale of lensing ``compact'' objects, because kinematical or velocity structures of lensing objects, even for DM in the MW halo region, are well constrained by various observations; the relative velocity of lens-source-observer determines the crossing time of the lensing Einstein radius, or equivalently the microlensing light curve timescale, once mass of a lensing object is assumed because 
the mass determines the Einstein radius. Thus we can use the invaluable OGLE data, which covers the wide range of timescales $t_{\rm E}\simeq [0.1,300]~{\rm days}$, allows us to explore the abundance of unknown ``compact''
objects over the wide range of mass scales. 

To do this, we first revisited the Galactic bulge/disk models to estimate event rates of microlensing due to astrophysical objects such as brown dwarfs, MS stars, and stellar remnants (white dwarfs, neutron stars, and astrophysical black holes), 
following Ref.~\cite{2017Natur.548..183M}. Since the mass of Galactic disk/bulge regions is dominated by low-mass main-sequence (MS) stars around $M\sim 0.5M_\odot$, astronomers have a quite good knowledge of the abundance of stars from various observations of star number counts. In addition, stellar remnants are from massive stars, so we can infer their abundances from the initial mass function
of low-mass MS stars that is well constrained by various observations. We showed that, 
even if details of our models would be different from that of Ref.~\cite{2017Natur.548..183M}, the standard Galactic bulge/disk models including MS stars and stellar remnants nicely reproduce the OGLE events at timescales $t_{\rm E}\simgt 20~{\rm days}$, corresponding to objects with $M\simgt 1M_\odot$. For the shorter-timescale events, we need to add contributions from brown dwarfs (BD) that are invisible or difficult to directly observe. The origin and nature of BDs ($0.01\simlt M/M_\odot\simlt 0.08$) are
not well understood. Some of BDs should form around a primary MS star, while some of BDs would form, as a primary gravitating object, in a protoplanetary disk. The BDs contributing the short timescale microlensing are ``unbounded'' BDs, because the timescale becomes too long if the host primary star contributes microlensing (because of much larger mass compared to that of BD). Nevertheless, the timescale distribution of OGLE data continuously extends to shorter timescales than the MS peak timescale, and it suggests a population of BD-mass objects with a continuous abundance to that of MS stars, except for the 6 ultrashort-timescale events in $t_{\rm E}\simeq [0.1,0.3]~{\rm days}$. We showed that our model can reproduce the entire timescale 
distribution for the main population of OGLE events if we adjust the abundance of BDs as done in \citet{2017Natur.548..183M} (about 0.18 BDs per MS star in our model). 

Given a justification of the standard Galactic bulge/disk models, we employ the ``null hypothesis'', i.e. no PBH microlensing event in the OGLE data, to obtain the stringent upper bound on the abundance of PBHs in the mass range $M_{\rm PBH}\simeq [10^{-6},10^{-3}]M_\odot$ (from Earth to Jupiter mass scales), assuming the monochromatic mass spectrum (Fig.~\ref{fig:upper_bound}). The upper bounds are tighter than the previous bound from the MACHO/EROS experiments \citep{EROS:07} and the Subaru/HSC microlensing search for M31 \citep{2017arXiv170102151N}. This result shows the power of microlensing for exploring the PBH abundance.

Even more interestingly, we showed that the 6 ultrashort-timescale events can be well explained by PBHs of Earth-mass scales if such PBHs constitute about 1\% of DM in the MW region (Fig.~\ref{fig:allowed_region}). Even if we employ the monochromatic mass spectrum for simplicity, the timescale distribution naturally arises from the velocity distribution of PBHs expected for the DM kinematical structures in the MW region. There is a mechanism in inflation model to produce PBHs in such a narrow mass range or with right abundance \citep[e.g.][]{2018PhRvD..97d3514I}. If this is a real PBH microlensing, 
it would be a big discover.  Such a small mass black hole cannot be made by any astrophysical process, so this would also give an evidence of the large primordial perturbations at the corresponding Hubble horizon in the early universe. If we include a possible distribution of PBH masses, the results would be changed. Nevertheless it is rather straightforward to translate our results into a specific PBH 
model with a given mass spectrum, following the methods developed in \citet{Carretal:16} \citep[also see][]{Inomataetal:17}.

A usual explanation of the ultrashort-timescale OGLE events is due to ``unbounded'' Earth-like planets, where ``unbounded'' is needed to have the right microlensing timescale. These unbounded, more exactly wide-orbit or free-floating planets, could be formed by the formation of planetary system or scattering of planetary systems; for example, if a planetary system encounters a massive planet or star, such an Earth-mass planet might be scattered. Even if this happens, why there is a gap around $t_{\rm E}\simeq 0.4~{\rm days}$ between the ultrashort-timescale events and the majority of events (i.e. the main population). Of course the gap might be an apparent statistical fluctuation due to the low number statistics, but it would be completely a mystery if 
if this is genuine, because there would be a continuous mass spectrum expected for  unbounded planets from Earth (or even smaller-masses) to Jupiter masses. These involve complicated, nonlinear astrophysics in planetary or star formation, so a further observational study would be a more direct path to resolving the origin and nature of these short timescale events.
For example, a more detailed study of lightcurve for each ultrashort-timescale event would be very useful  \citep[see][for such an attempt]{2018arXiv181100441M}. This requires a denser cadence data of microlensing search. Such a denser cadence data would be also useful to distinguish genuine microlensing events from other contaminating events such as stellar flare. 

To confirm or falsify the PBH hypothesis of short-timescale events against free-floating planets, there is a very promising, robust way. It is a microlensing search for stars in M31 using the Subaru HSC data (or eventually LSST data towards the Magellanic Clouds). 
The angular direction of M31 is in a high latitude in the Galactic coordinates, i.e. far from the Galactic disk. If the short timescale microlensing events are due to free-floating planets, we should expect a much smaller number of events towards M31, because there is a much less number of stars in the high latitude direction compared to the direction to the Galactic center. On the other hand, if the PBH scenario is true, we should find microlensing events with a frequency predicted by the standard 
DM model of the MW halo region that is supported by the disk rotation measurements.  As shown in \citet{2017arXiv170102151N}, the Subaru and HSC combination is ideal because its FoV can cover the entire disk region of M31 and the large aperture allows us to use main-sequence stars in M31 for the microlensing search even with short exposure (e.g., 90~sec). We are now carrying out a monitoring observation of M31 with Subaru HSC, and we envision that we can address these important questions in the near future.

\section*{Acknowledgments}
We would like to thank the OGLE collaboration for making their microlensing data publicly available to us via \citet{2017Natur.548..183M}. We also thank to Chiaki Kobayashi, Toshiki Kurita, Hitoshi Murayama, Toyokazu Sekiguchi, Sunao Sugiyama and Misao Sasaki for useful discussion. 
This research was supported by World Premier International Research Center Initiative (WPI), MEXT, Japan.
This work was in part supported by MEXT Grant-in-Aid for Scientific Research on Innovative Areas (No.~JP15H05887, JP15H05893, JP15K21733, JP18H04356), and by JSPS KAKENHI Grant Numbers JP17J03653 (HN), JP15H03654 (MT), and 15K17659 (SY).


\bibliography{refs}

\begin{thebibliography}{62}
\expandafter\ifx\csname natexlab\endcsname\relax\def\natexlab#1{#1}\fi
\expandafter\ifx\csname bibnamefont\endcsname\relax
  \def\bibnamefont#1{#1}\fi
\expandafter\ifx\csname bibfnamefont\endcsname\relax
  \def\bibfnamefont#1{#1}\fi
\expandafter\ifx\csname citenamefont\endcsname\relax
  \def\citenamefont#1{#1}\fi
\expandafter\ifx\csname url\endcsname\relax
  \def\url#1{\texttt{#1}}\fi
\expandafter\ifx\csname urlprefix\endcsname\relax\def\urlprefix{URL }\fi
\providecommand{\bibinfo}[2]{#2}
\providecommand{\eprint}[2][]{\url{#2}}

\bibitem[{\citenamefont{{Springel} et~al.}(2005)\citenamefont{{Springel},
  {White}, {Jenkins}, {Frenk}, {Yoshida}, {Gao}, {Navarro}, {Thacker},
  {Croton}, {Helly} et~al.}}]{2005Natur.435..629S}
\bibinfo{author}{\bibfnamefont{V.}~\bibnamefont{{Springel}}},
  \bibinfo{author}{\bibfnamefont{S.~D.~M.} \bibnamefont{{White}}},
  \bibinfo{author}{\bibfnamefont{A.}~\bibnamefont{{Jenkins}}},
  \bibinfo{author}{\bibfnamefont{C.~S.} \bibnamefont{{Frenk}}},
  \bibinfo{author}{\bibfnamefont{N.}~\bibnamefont{{Yoshida}}},
  \bibinfo{author}{\bibfnamefont{L.}~\bibnamefont{{Gao}}},
  \bibinfo{author}{\bibfnamefont{J.}~\bibnamefont{{Navarro}}},
  \bibinfo{author}{\bibfnamefont{R.}~\bibnamefont{{Thacker}}},
  \bibinfo{author}{\bibfnamefont{D.}~\bibnamefont{{Croton}}},
  \bibinfo{author}{\bibfnamefont{J.}~\bibnamefont{{Helly}}},
  \bibnamefont{et~al.}, \bibinfo{journal}{\nat} \textbf{\bibinfo{volume}{435}},
  \bibinfo{pages}{629} (\bibinfo{year}{2005}), \eprint{astro-ph/0504097}.

\bibitem[{\citenamefont{{Jungman} et~al.}(1996)\citenamefont{{Jungman},
  {Kamionkowski}, and {Griest}}}]{Jungmanetal:96}
\bibinfo{author}{\bibfnamefont{G.}~\bibnamefont{{Jungman}}},
  \bibinfo{author}{\bibfnamefont{M.}~\bibnamefont{{Kamionkowski}}},
  \bibnamefont{and} \bibinfo{author}{\bibfnamefont{K.}~\bibnamefont{{Griest}}},
  \bibinfo{journal}{\physrep} \textbf{\bibinfo{volume}{267}},
  \bibinfo{pages}{195} (\bibinfo{year}{1996}), \eprint{hep-ph/9506380}.

\bibitem[{\citenamefont{Arina}(2018)}]{Arina:2018zcq}
\bibinfo{author}{\bibfnamefont{C.}~\bibnamefont{Arina}} (\bibinfo{year}{2018}),
  \eprint{1805.04290}.

\bibitem[{\citenamefont{Hooper}(2018)}]{Hooper:2018kfv}
\bibinfo{author}{\bibfnamefont{D.}~\bibnamefont{Hooper}}
  (\bibinfo{year}{2018}), \eprint{1812.02029}.

\bibitem[{\citenamefont{{Hawking}}(1971)}]{Hawking:71}
\bibinfo{author}{\bibfnamefont{S.}~\bibnamefont{{Hawking}}},
  \bibinfo{journal}{\mnras} \textbf{\bibinfo{volume}{152}}, \bibinfo{pages}{75}
  (\bibinfo{year}{1971}).

\bibitem[{\citenamefont{{Carr} and {Hawking}}(1974)}]{CarrHawking:74}
\bibinfo{author}{\bibfnamefont{B.~J.} \bibnamefont{{Carr}}} \bibnamefont{and}
  \bibinfo{author}{\bibfnamefont{S.~W.} \bibnamefont{{Hawking}}},
  \bibinfo{journal}{\mnras} \textbf{\bibinfo{volume}{168}},
  \bibinfo{pages}{399} (\bibinfo{year}{1974}).

\bibitem[{\citenamefont{{Carr}}(1975)}]{Carr:75}
\bibinfo{author}{\bibfnamefont{B.~J.} \bibnamefont{{Carr}}},
  \bibinfo{journal}{\apj} \textbf{\bibinfo{volume}{201}}, \bibinfo{pages}{1}
  (\bibinfo{year}{1975}).

\bibitem[{\citenamefont{{Carr} et~al.}(2016)\citenamefont{{Carr}, {K{\"u}hnel},
  and {Sandstad}}}]{Carretal:16}
\bibinfo{author}{\bibfnamefont{B.}~\bibnamefont{{Carr}}},
  \bibinfo{author}{\bibfnamefont{F.}~\bibnamefont{{K{\"u}hnel}}},
  \bibnamefont{and}
  \bibinfo{author}{\bibfnamefont{M.}~\bibnamefont{{Sandstad}}},
  \bibinfo{journal}{\prd} \textbf{\bibinfo{volume}{94}}, \bibinfo{eid}{083504}
  (\bibinfo{year}{2016}), \eprint{1607.06077}.

\bibitem[{\citenamefont{{Inomata}
  et~al.}(2017{\natexlab{a}})\citenamefont{{Inomata}, {Kawasaki}, {Mukaida},
  {Tada}, and {Yanagida}}}]{Inomataetal:16}
\bibinfo{author}{\bibfnamefont{K.}~\bibnamefont{{Inomata}}},
  \bibinfo{author}{\bibfnamefont{M.}~\bibnamefont{{Kawasaki}}},
  \bibinfo{author}{\bibfnamefont{K.}~\bibnamefont{{Mukaida}}},
  \bibinfo{author}{\bibfnamefont{Y.}~\bibnamefont{{Tada}}}, \bibnamefont{and}
  \bibinfo{author}{\bibfnamefont{T.~T.} \bibnamefont{{Yanagida}}},
  \bibinfo{journal}{\prd} \textbf{\bibinfo{volume}{95}}, \bibinfo{eid}{123510}
  (\bibinfo{year}{2017}{\natexlab{a}}), \eprint{1611.06130}.

\bibitem[{\citenamefont{{Cotner} and {Kusenko}}(2017)}]{2017PhRvL.119c1103C}
\bibinfo{author}{\bibfnamefont{E.}~\bibnamefont{{Cotner}}} \bibnamefont{and}
  \bibinfo{author}{\bibfnamefont{A.}~\bibnamefont{{Kusenko}}},
  \bibinfo{journal}{\prl} \textbf{\bibinfo{volume}{119}}, \bibinfo{eid}{031103}
  (\bibinfo{year}{2017}), \eprint{1612.02529}.

\bibitem[{\citenamefont{{Cotner} et~al.}(2018)\citenamefont{{Cotner},
  {Kusenko}, and {Takhistov}}}]{2018PhRvD..98h3513C}
\bibinfo{author}{\bibfnamefont{E.}~\bibnamefont{{Cotner}}},
  \bibinfo{author}{\bibfnamefont{A.}~\bibnamefont{{Kusenko}}},
  \bibnamefont{and}
  \bibinfo{author}{\bibfnamefont{V.}~\bibnamefont{{Takhistov}}},
  \bibinfo{journal}{\prd} \textbf{\bibinfo{volume}{98}}, \bibinfo{eid}{083513}
  (\bibinfo{year}{2018}), \eprint{1801.03321}.

\bibitem[{\citenamefont{{Sasaki} et~al.}(2016)\citenamefont{{Sasaki}, {Suyama},
  {Tanaka}, and {Yokoyama}}}]{Sasakietal:16}
\bibinfo{author}{\bibfnamefont{M.}~\bibnamefont{{Sasaki}}},
  \bibinfo{author}{\bibfnamefont{T.}~\bibnamefont{{Suyama}}},
  \bibinfo{author}{\bibfnamefont{T.}~\bibnamefont{{Tanaka}}}, \bibnamefont{and}
  \bibinfo{author}{\bibfnamefont{S.}~\bibnamefont{{Yokoyama}}},
  \bibinfo{journal}{Physical Review Letters} \textbf{\bibinfo{volume}{117}},
  \bibinfo{eid}{061101} (\bibinfo{year}{2016}), \eprint{1603.08338}.

\bibitem[{\citenamefont{{Bird} et~al.}(2016)\citenamefont{{Bird}, {Cholis},
  {Mu{\~n}oz}, {Ali-Ha{\"i}moud}, {Kamionkowski}, {Kovetz}, {Raccanelli}, and
  {Riess}}}]{Birdetal:16}
\bibinfo{author}{\bibfnamefont{S.}~\bibnamefont{{Bird}}},
  \bibinfo{author}{\bibfnamefont{I.}~\bibnamefont{{Cholis}}},
  \bibinfo{author}{\bibfnamefont{J.~B.} \bibnamefont{{Mu{\~n}oz}}},
  \bibinfo{author}{\bibfnamefont{Y.}~\bibnamefont{{Ali-Ha{\"i}moud}}},
  \bibinfo{author}{\bibfnamefont{M.}~\bibnamefont{{Kamionkowski}}},
  \bibinfo{author}{\bibfnamefont{E.~D.} \bibnamefont{{Kovetz}}},
  \bibinfo{author}{\bibfnamefont{A.}~\bibnamefont{{Raccanelli}}},
  \bibnamefont{and} \bibinfo{author}{\bibfnamefont{A.~G.}
  \bibnamefont{{Riess}}}, \bibinfo{journal}{Physical Review Letters}
  \textbf{\bibinfo{volume}{116}}, \bibinfo{eid}{201301} (\bibinfo{year}{2016}),
  \eprint{1603.00464}.

\bibitem[{\citenamefont{{Sasaki} et~al.}(2018)\citenamefont{{Sasaki}, {Suyama},
  {Tanaka}, and {Yokoyama}}}]{2018CQGra..35f3001S}
\bibinfo{author}{\bibfnamefont{M.}~\bibnamefont{{Sasaki}}},
  \bibinfo{author}{\bibfnamefont{T.}~\bibnamefont{{Suyama}}},
  \bibinfo{author}{\bibfnamefont{T.}~\bibnamefont{{Tanaka}}}, \bibnamefont{and}
  \bibinfo{author}{\bibfnamefont{S.}~\bibnamefont{{Yokoyama}}},
  \bibinfo{journal}{Classical and Quantum Gravity}
  \textbf{\bibinfo{volume}{35}}, \bibinfo{eid}{063001} (\bibinfo{year}{2018}),
  \eprint{1801.05235}.

\bibitem[{\citenamefont{{Ali-Ha{\"i}moud}
  et~al.}(2017)\citenamefont{{Ali-Ha{\"i}moud}, {Kovetz}, and
  {Kamionkowski}}}]{2017PhRvD..96l3523A}
\bibinfo{author}{\bibfnamefont{Y.}~\bibnamefont{{Ali-Ha{\"i}moud}}},
  \bibinfo{author}{\bibfnamefont{E.~D.} \bibnamefont{{Kovetz}}},
  \bibnamefont{and}
  \bibinfo{author}{\bibfnamefont{M.}~\bibnamefont{{Kamionkowski}}},
  \bibinfo{journal}{\prd} \textbf{\bibinfo{volume}{96}}, \bibinfo{eid}{123523}
  (\bibinfo{year}{2017}), \eprint{1709.06576}.

\bibitem[{\citenamefont{{Nakamura} et~al.}(2016)\citenamefont{{Nakamura},
  {Ando}, {Kinugawa}, {Nakano}, {Eda}, {Sato}, {Musha}, {Akutsu}, {Tanaka},
  {Seto} et~al.}}]{2016PTEP.2016i3E01N}
\bibinfo{author}{\bibfnamefont{T.}~\bibnamefont{{Nakamura}}},
  \bibinfo{author}{\bibfnamefont{M.}~\bibnamefont{{Ando}}},
  \bibinfo{author}{\bibfnamefont{T.}~\bibnamefont{{Kinugawa}}},
  \bibinfo{author}{\bibfnamefont{H.}~\bibnamefont{{Nakano}}},
  \bibinfo{author}{\bibfnamefont{K.}~\bibnamefont{{Eda}}},
  \bibinfo{author}{\bibfnamefont{S.}~\bibnamefont{{Sato}}},
  \bibinfo{author}{\bibfnamefont{M.}~\bibnamefont{{Musha}}},
  \bibinfo{author}{\bibfnamefont{T.}~\bibnamefont{{Akutsu}}},
  \bibinfo{author}{\bibfnamefont{T.}~\bibnamefont{{Tanaka}}},
  \bibinfo{author}{\bibfnamefont{N.}~\bibnamefont{{Seto}}},
  \bibnamefont{et~al.}, \bibinfo{journal}{Progress of Theoretical and
  Experimental Physics} \textbf{\bibinfo{volume}{2016}}, \bibinfo{eid}{093E01}
  (\bibinfo{year}{2016}), \eprint{1607.00897}.

\bibitem[{\citenamefont{{Ioka} et~al.}(1998)\citenamefont{{Ioka}, {Chiba},
  {Tanaka}, and {Nakamura}}}]{1998PhRvD..58f3003I}
\bibinfo{author}{\bibfnamefont{K.}~\bibnamefont{{Ioka}}},
  \bibinfo{author}{\bibfnamefont{T.}~\bibnamefont{{Chiba}}},
  \bibinfo{author}{\bibfnamefont{T.}~\bibnamefont{{Tanaka}}}, \bibnamefont{and}
  \bibinfo{author}{\bibfnamefont{T.}~\bibnamefont{{Nakamura}}},
  \bibinfo{journal}{\prd} \textbf{\bibinfo{volume}{58}}, \bibinfo{eid}{063003}
  (\bibinfo{year}{1998}), \eprint{astro-ph/9807018}.

\bibitem[{\citenamefont{{Abbott} et~al.}(2016)\citenamefont{{Abbott}, {Abbott},
  {Abbott}, {Abernathy}, {Acernese}, {Ackley}, {Adams}, {Adams}, {Addesso},
  {Adhikari} et~al.}}]{LIGO:16}
\bibinfo{author}{\bibfnamefont{B.~P.} \bibnamefont{{Abbott}}},
  \bibinfo{author}{\bibfnamefont{R.}~\bibnamefont{{Abbott}}},
  \bibinfo{author}{\bibfnamefont{T.~D.} \bibnamefont{{Abbott}}},
  \bibinfo{author}{\bibfnamefont{M.~R.} \bibnamefont{{Abernathy}}},
  \bibinfo{author}{\bibfnamefont{F.}~\bibnamefont{{Acernese}}},
  \bibinfo{author}{\bibfnamefont{K.}~\bibnamefont{{Ackley}}},
  \bibinfo{author}{\bibfnamefont{C.}~\bibnamefont{{Adams}}},
  \bibinfo{author}{\bibfnamefont{T.}~\bibnamefont{{Adams}}},
  \bibinfo{author}{\bibfnamefont{P.}~\bibnamefont{{Addesso}}},
  \bibinfo{author}{\bibfnamefont{R.~X.} \bibnamefont{{Adhikari}}},
  \bibnamefont{et~al.}, \bibinfo{journal}{Physical Review Letters}
  \textbf{\bibinfo{volume}{116}}, \bibinfo{eid}{061102} (\bibinfo{year}{2016}),
  \eprint{1602.03837}.

\bibitem[{\citenamefont{Abbott and the others}(2017)\citenamefont{Abbott
  et~al.}}]{PhysRevLett.119.141101}
\bibinfo{author}{\bibfnamefont{B.~P.} \bibnamefont{Abbott}}
  \bibnamefont{et~al.} (\bibinfo{collaboration}{LIGO Scientific Collaboration
  and Virgo Collaboration}), \bibinfo{journal}{Phys. Rev. Lett.}
  \textbf{\bibinfo{volume}{119}}, \bibinfo{pages}{141101}
  (\bibinfo{year}{2017}),
  \urlprefix\url{https://link.aps.org/doi/10.1103/PhysRevLett.119.141101}.

\bibitem[{\citenamefont{{Carr} et~al.}(2010)\citenamefont{{Carr}, {Kohri},
  {Sendouda}, and {Yokoyama}}}]{Carretal:10}
\bibinfo{author}{\bibfnamefont{B.~J.} \bibnamefont{{Carr}}},
  \bibinfo{author}{\bibfnamefont{K.}~\bibnamefont{{Kohri}}},
  \bibinfo{author}{\bibfnamefont{Y.}~\bibnamefont{{Sendouda}}},
  \bibnamefont{and}
  \bibinfo{author}{\bibfnamefont{J.}~\bibnamefont{{Yokoyama}}},
  \bibinfo{journal}{\prd} \textbf{\bibinfo{volume}{81}}, \bibinfo{eid}{104019}
  (\bibinfo{year}{2010}), \eprint{0912.5297}.

\bibitem[{\citenamefont{{Gould}}(1992)}]{Gould:92}
\bibinfo{author}{\bibfnamefont{A.}~\bibnamefont{{Gould}}},
  \bibinfo{journal}{\apjl} \textbf{\bibinfo{volume}{386}}, \bibinfo{pages}{L5}
  (\bibinfo{year}{1992}).

\bibitem[{\citenamefont{{Katz} et~al.}(2018)\citenamefont{{Katz}, {Kopp},
  {Sibiryakov}, and {Xue}}}]{2018JCAP...12..005K}
\bibinfo{author}{\bibfnamefont{A.}~\bibnamefont{{Katz}}},
  \bibinfo{author}{\bibfnamefont{J.}~\bibnamefont{{Kopp}}},
  \bibinfo{author}{\bibfnamefont{S.}~\bibnamefont{{Sibiryakov}}},
  \bibnamefont{and} \bibinfo{author}{\bibfnamefont{W.}~\bibnamefont{{Xue}}},
  \bibinfo{journal}{\jcap} \textbf{\bibinfo{volume}{12}}, \bibinfo{eid}{005}
  (\bibinfo{year}{2018}), \eprint{1807.11495}.

\bibitem[{\citenamefont{{Graham} et~al.}(2015)\citenamefont{{Graham},
  {Rajendran}, and {Varela}}}]{2015PhRvD..92f3007G}
\bibinfo{author}{\bibfnamefont{P.~W.} \bibnamefont{{Graham}}},
  \bibinfo{author}{\bibfnamefont{S.}~\bibnamefont{{Rajendran}}},
  \bibnamefont{and} \bibinfo{author}{\bibfnamefont{J.}~\bibnamefont{{Varela}}},
  \bibinfo{journal}{\prd} \textbf{\bibinfo{volume}{92}}, \bibinfo{eid}{063007}
  (\bibinfo{year}{2015}), \eprint{1505.04444}.

\bibitem[{\citenamefont{{Capela} et~al.}(2013)\citenamefont{{Capela},
  {Pshirkov}, and {Tinyakov}}}]{Capelaetal:13b}
\bibinfo{author}{\bibfnamefont{F.}~\bibnamefont{{Capela}}},
  \bibinfo{author}{\bibfnamefont{M.}~\bibnamefont{{Pshirkov}}},
  \bibnamefont{and}
  \bibinfo{author}{\bibfnamefont{P.}~\bibnamefont{{Tinyakov}}},
  \bibinfo{journal}{\prd} \textbf{\bibinfo{volume}{87}}, \bibinfo{eid}{123524}
  (\bibinfo{year}{2013}), \eprint{1301.4984}.

\bibitem[{\citenamefont{{Fuller} et~al.}(2017)\citenamefont{{Fuller},
  {Kusenko}, and {Takhistov}}}]{2017PhRvL.119f1101F}
\bibinfo{author}{\bibfnamefont{G.~M.} \bibnamefont{{Fuller}}},
  \bibinfo{author}{\bibfnamefont{A.}~\bibnamefont{{Kusenko}}},
  \bibnamefont{and}
  \bibinfo{author}{\bibfnamefont{V.}~\bibnamefont{{Takhistov}}},
  \bibinfo{journal}{Physical Review Letters} \textbf{\bibinfo{volume}{119}},
  \bibinfo{eid}{061101} (\bibinfo{year}{2017}), \eprint{1704.01129}.

\bibitem[{\citenamefont{{Alcock} et~al.}(2000)\citenamefont{{Alcock},
  {Allsman}, {Alves}, {Axelrod}, {Becker}, {Bennett}, {Cook}, {Dalal}, {Drake},
  {Freeman} et~al.}}]{Alcocketal:00}
\bibinfo{author}{\bibfnamefont{C.}~\bibnamefont{{Alcock}}},
  \bibinfo{author}{\bibfnamefont{R.~A.} \bibnamefont{{Allsman}}},
  \bibinfo{author}{\bibfnamefont{D.~R.} \bibnamefont{{Alves}}},
  \bibinfo{author}{\bibfnamefont{T.~S.} \bibnamefont{{Axelrod}}},
  \bibinfo{author}{\bibfnamefont{A.~C.} \bibnamefont{{Becker}}},
  \bibinfo{author}{\bibfnamefont{D.~P.} \bibnamefont{{Bennett}}},
  \bibinfo{author}{\bibfnamefont{K.~H.} \bibnamefont{{Cook}}},
  \bibinfo{author}{\bibfnamefont{N.}~\bibnamefont{{Dalal}}},
  \bibinfo{author}{\bibfnamefont{A.~J.} \bibnamefont{{Drake}}},
  \bibinfo{author}{\bibfnamefont{K.~C.} \bibnamefont{{Freeman}}},
  \bibnamefont{et~al.}, \bibinfo{journal}{\apj} \textbf{\bibinfo{volume}{542}},
  \bibinfo{pages}{281} (\bibinfo{year}{2000}), \eprint{astro-ph/0001272}.

\bibitem[{\citenamefont{{Tisserand} et~al.}(2007)\citenamefont{{Tisserand}, {Le
  Guillou}, {Afonso}, {Albert}, {Andersen}, {Ansari}, {Aubourg}, {Bareyre},
  {Beaulieu}, {Charlot} et~al.}}]{EROS:07}
\bibinfo{author}{\bibfnamefont{P.}~\bibnamefont{{Tisserand}}},
  \bibinfo{author}{\bibfnamefont{L.}~\bibnamefont{{Le Guillou}}},
  \bibinfo{author}{\bibfnamefont{C.}~\bibnamefont{{Afonso}}},
  \bibinfo{author}{\bibfnamefont{J.~N.} \bibnamefont{{Albert}}},
  \bibinfo{author}{\bibfnamefont{J.}~\bibnamefont{{Andersen}}},
  \bibinfo{author}{\bibfnamefont{R.}~\bibnamefont{{Ansari}}},
  \bibinfo{author}{\bibfnamefont{{\'E}.}~\bibnamefont{{Aubourg}}},
  \bibinfo{author}{\bibfnamefont{P.}~\bibnamefont{{Bareyre}}},
  \bibinfo{author}{\bibfnamefont{J.~P.} \bibnamefont{{Beaulieu}}},
  \bibinfo{author}{\bibfnamefont{X.}~\bibnamefont{{Charlot}}},
  \bibnamefont{et~al.}, \bibinfo{journal}{\aap} \textbf{\bibinfo{volume}{469}},
  \bibinfo{pages}{387} (\bibinfo{year}{2007}), \eprint{astro-ph/0607207}.

\bibitem[{\citenamefont{{Griest} et~al.}(2014)\citenamefont{{Griest},
  {Cieplak}, and {Lehner}}}]{Griestetal:14}
\bibinfo{author}{\bibfnamefont{K.}~\bibnamefont{{Griest}}},
  \bibinfo{author}{\bibfnamefont{A.~M.} \bibnamefont{{Cieplak}}},
  \bibnamefont{and} \bibinfo{author}{\bibfnamefont{M.~J.}
  \bibnamefont{{Lehner}}}, \bibinfo{journal}{\apj}
  \textbf{\bibinfo{volume}{786}}, \bibinfo{eid}{158} (\bibinfo{year}{2014}),
  \eprint{1307.5798}.

\bibitem[{\citenamefont{{Niikura} et~al.}(2017)\citenamefont{{Niikura},
  {Takada}, {Yasuda}, {Lupton}, {Sumi}, {More}, {Kurita}, {Sugiyama}, {More},
  {Oguri} et~al.}}]{2017arXiv170102151N}
\bibinfo{author}{\bibfnamefont{H.}~\bibnamefont{{Niikura}}},
  \bibinfo{author}{\bibfnamefont{M.}~\bibnamefont{{Takada}}},
  \bibinfo{author}{\bibfnamefont{N.}~\bibnamefont{{Yasuda}}},
  \bibinfo{author}{\bibfnamefont{R.~H.} \bibnamefont{{Lupton}}},
  \bibinfo{author}{\bibfnamefont{T.}~\bibnamefont{{Sumi}}},
  \bibinfo{author}{\bibfnamefont{S.}~\bibnamefont{{More}}},
  \bibinfo{author}{\bibfnamefont{T.}~\bibnamefont{{Kurita}}},
  \bibinfo{author}{\bibfnamefont{S.}~\bibnamefont{{Sugiyama}}},
  \bibinfo{author}{\bibfnamefont{A.}~\bibnamefont{{More}}},
  \bibinfo{author}{\bibfnamefont{M.}~\bibnamefont{{Oguri}}},
  \bibnamefont{et~al.}, \bibinfo{journal}{ArXiv e-prints}
  (\bibinfo{year}{2017}), \eprint{1701.02151}.

\bibitem[{\citenamefont{{Kelly} et~al.}(2018)\citenamefont{{Kelly}, {Diego},
  {Rodney}, {Kaiser}, {Broadhurst}, {Zitrin}, {Treu}, {P{\'e}rez-Gonz{\'a}lez},
  {Morishita}, {Jauzac} et~al.}}]{2018NatAs...2..334K}
\bibinfo{author}{\bibfnamefont{P.~L.} \bibnamefont{{Kelly}}},
  \bibinfo{author}{\bibfnamefont{J.~M.} \bibnamefont{{Diego}}},
  \bibinfo{author}{\bibfnamefont{S.}~\bibnamefont{{Rodney}}},
  \bibinfo{author}{\bibfnamefont{N.}~\bibnamefont{{Kaiser}}},
  \bibinfo{author}{\bibfnamefont{T.}~\bibnamefont{{Broadhurst}}},
  \bibinfo{author}{\bibfnamefont{A.}~\bibnamefont{{Zitrin}}},
  \bibinfo{author}{\bibfnamefont{T.}~\bibnamefont{{Treu}}},
  \bibinfo{author}{\bibfnamefont{P.~G.}
  \bibnamefont{{P{\'e}rez-Gonz{\'a}lez}}},
  \bibinfo{author}{\bibfnamefont{T.}~\bibnamefont{{Morishita}}},
  \bibinfo{author}{\bibfnamefont{M.}~\bibnamefont{{Jauzac}}},
  \bibnamefont{et~al.}, \bibinfo{journal}{Nature Astronomy}
  \textbf{\bibinfo{volume}{2}}, \bibinfo{pages}{334} (\bibinfo{year}{2018}),
  \eprint{1706.10279}.

\bibitem[{\citenamefont{{Diego} et~al.}(2018)\citenamefont{{Diego}, {Kaiser},
  {Broadhurst}, {Kelly}, {Rodney}, {Morishita}, {Oguri}, {Ross}, {Zitrin},
  {Jauzac} et~al.}}]{2018ApJ...857...25D}
\bibinfo{author}{\bibfnamefont{J.~M.} \bibnamefont{{Diego}}},
  \bibinfo{author}{\bibfnamefont{N.}~\bibnamefont{{Kaiser}}},
  \bibinfo{author}{\bibfnamefont{T.}~\bibnamefont{{Broadhurst}}},
  \bibinfo{author}{\bibfnamefont{P.~L.} \bibnamefont{{Kelly}}},
  \bibinfo{author}{\bibfnamefont{S.}~\bibnamefont{{Rodney}}},
  \bibinfo{author}{\bibfnamefont{T.}~\bibnamefont{{Morishita}}},
  \bibinfo{author}{\bibfnamefont{M.}~\bibnamefont{{Oguri}}},
  \bibinfo{author}{\bibfnamefont{T.~W.} \bibnamefont{{Ross}}},
  \bibinfo{author}{\bibfnamefont{A.}~\bibnamefont{{Zitrin}}},
  \bibinfo{author}{\bibfnamefont{M.}~\bibnamefont{{Jauzac}}},
  \bibnamefont{et~al.}, \bibinfo{journal}{\apj} \textbf{\bibinfo{volume}{857}},
  \bibinfo{eid}{25} (\bibinfo{year}{2018}), \eprint{1706.10281}.

\bibitem[{\citenamefont{{Oguri} et~al.}(2018)\citenamefont{{Oguri}, {Diego},
  {Kaiser}, {Kelly}, and {Broadhurst}}}]{2018PhRvD..97b3518O}
\bibinfo{author}{\bibfnamefont{M.}~\bibnamefont{{Oguri}}},
  \bibinfo{author}{\bibfnamefont{J.~M.} \bibnamefont{{Diego}}},
  \bibinfo{author}{\bibfnamefont{N.}~\bibnamefont{{Kaiser}}},
  \bibinfo{author}{\bibfnamefont{P.~L.} \bibnamefont{{Kelly}}},
  \bibnamefont{and}
  \bibinfo{author}{\bibfnamefont{T.}~\bibnamefont{{Broadhurst}}},
  \bibinfo{journal}{\prd} \textbf{\bibinfo{volume}{97}}, \bibinfo{eid}{023518}
  (\bibinfo{year}{2018}), \eprint{1710.00148}.

\bibitem[{\citenamefont{{Inoue} and {Kusenko}}(2017)}]{2017JCAP...10..034I}
\bibinfo{author}{\bibfnamefont{Y.}~\bibnamefont{{Inoue}}} \bibnamefont{and}
  \bibinfo{author}{\bibfnamefont{A.}~\bibnamefont{{Kusenko}}},
  \bibinfo{journal}{Journal of Cosmology and Astro-Particle Physics}
  \textbf{\bibinfo{volume}{2017}}, \bibinfo{eid}{034} (\bibinfo{year}{2017}),
  \eprint{1705.00791}.

\bibitem[{\citenamefont{{Ricotti} et~al.}(2008)\citenamefont{{Ricotti},
  {Ostriker}, and {Mack}}}]{Ricottietal:08}
\bibinfo{author}{\bibfnamefont{M.}~\bibnamefont{{Ricotti}}},
  \bibinfo{author}{\bibfnamefont{J.~P.} \bibnamefont{{Ostriker}}},
  \bibnamefont{and} \bibinfo{author}{\bibfnamefont{K.~J.}
  \bibnamefont{{Mack}}}, \bibinfo{journal}{\apj}
  \textbf{\bibinfo{volume}{680}}, \bibinfo{eid}{829-845}
  (\bibinfo{year}{2008}), \eprint{0709.0524}.

\bibitem[{\citenamefont{{Ali-Ha{\"i}moud} and
  {Kamionkowski}}(2017)}]{Ali-HaimoudKamionkowski:17}
\bibinfo{author}{\bibfnamefont{Y.}~\bibnamefont{{Ali-Ha{\"i}moud}}}
  \bibnamefont{and}
  \bibinfo{author}{\bibfnamefont{M.}~\bibnamefont{{Kamionkowski}}},
  \bibinfo{journal}{\prd} \textbf{\bibinfo{volume}{95}}, \bibinfo{eid}{043534}
  (\bibinfo{year}{2017}), \eprint{1612.05644}.

\bibitem[{\citenamefont{{Aloni} et~al.}(2017)\citenamefont{{Aloni}, {Blum}, and
  {Flauger}}}]{2017JCAP...05..017A}
\bibinfo{author}{\bibfnamefont{D.}~\bibnamefont{{Aloni}}},
  \bibinfo{author}{\bibfnamefont{K.}~\bibnamefont{{Blum}}}, \bibnamefont{and}
  \bibinfo{author}{\bibfnamefont{R.}~\bibnamefont{{Flauger}}},
  \bibinfo{journal}{\jcap} \textbf{\bibinfo{volume}{5}}, \bibinfo{eid}{017}
  (\bibinfo{year}{2017}), \eprint{1612.06811}.

\bibitem[{\citenamefont{{Schutz} and {Liu}}(2017)}]{2017PhRvD..95b3002S}
\bibinfo{author}{\bibfnamefont{K.}~\bibnamefont{{Schutz}}} \bibnamefont{and}
  \bibinfo{author}{\bibfnamefont{A.}~\bibnamefont{{Liu}}},
  \bibinfo{journal}{\prd} \textbf{\bibinfo{volume}{95}}, \bibinfo{eid}{023002}
  (\bibinfo{year}{2017}), \eprint{1610.04234}.

\bibitem[{\citenamefont{{Zumalac{\'a}rregui} and
  {Seljak}}(2018)}]{2018PhRvL.121n1101Z}
\bibinfo{author}{\bibfnamefont{M.}~\bibnamefont{{Zumalac{\'a}rregui}}}
  \bibnamefont{and} \bibinfo{author}{\bibfnamefont{U.}~\bibnamefont{{Seljak}}},
  \bibinfo{journal}{Physical Review Letters} \textbf{\bibinfo{volume}{121}},
  \bibinfo{eid}{141101} (\bibinfo{year}{2018}), \eprint{1712.02240}.

\bibitem[{\citenamefont{{Paczynski}}(1986)}]{Paczynski:86}
\bibinfo{author}{\bibfnamefont{B.}~\bibnamefont{{Paczynski}}},
  \bibinfo{journal}{\apj} \textbf{\bibinfo{volume}{304}}, \bibinfo{pages}{1}
  (\bibinfo{year}{1986}).

\bibitem[{\citenamefont{{Griest} et~al.}(1991)\citenamefont{{Griest}, {Alcock},
  {Axelrod}, {Bennett}, {Cook}, {Freeman}, {Park}, {Perlmutter}, {Peterson},
  {Quinn} et~al.}}]{Griestetal:91}
\bibinfo{author}{\bibfnamefont{K.}~\bibnamefont{{Griest}}},
  \bibinfo{author}{\bibfnamefont{C.}~\bibnamefont{{Alcock}}},
  \bibinfo{author}{\bibfnamefont{T.~S.} \bibnamefont{{Axelrod}}},
  \bibinfo{author}{\bibfnamefont{D.~P.} \bibnamefont{{Bennett}}},
  \bibinfo{author}{\bibfnamefont{K.~H.} \bibnamefont{{Cook}}},
  \bibinfo{author}{\bibfnamefont{K.~C.} \bibnamefont{{Freeman}}},
  \bibinfo{author}{\bibfnamefont{H.-S.} \bibnamefont{{Park}}},
  \bibinfo{author}{\bibfnamefont{S.}~\bibnamefont{{Perlmutter}}},
  \bibinfo{author}{\bibfnamefont{B.~A.} \bibnamefont{{Peterson}}},
  \bibinfo{author}{\bibfnamefont{P.~J.} \bibnamefont{{Quinn}}},
  \bibnamefont{et~al.}, \bibinfo{journal}{\apjl}
  \textbf{\bibinfo{volume}{372}}, \bibinfo{pages}{L79} (\bibinfo{year}{1991}).

\bibitem[{\citenamefont{{Udalski} et~al.}(1994)\citenamefont{{Udalski},
  {Szymanski}, {Kaluzny}, {Kubiak}, {Mateo}, and
  {Krzeminski}}}]{1994ApJ...426L..69U}
\bibinfo{author}{\bibfnamefont{A.}~\bibnamefont{{Udalski}}},
  \bibinfo{author}{\bibfnamefont{M.}~\bibnamefont{{Szymanski}}},
  \bibinfo{author}{\bibfnamefont{J.}~\bibnamefont{{Kaluzny}}},
  \bibinfo{author}{\bibfnamefont{M.}~\bibnamefont{{Kubiak}}},
  \bibinfo{author}{\bibfnamefont{M.}~\bibnamefont{{Mateo}}}, \bibnamefont{and}
  \bibinfo{author}{\bibfnamefont{W.}~\bibnamefont{{Krzeminski}}},
  \bibinfo{journal}{\apjl} \textbf{\bibinfo{volume}{426}}, \bibinfo{pages}{69}
  (\bibinfo{year}{1994}).

\bibitem[{\citenamefont{{Udalski} et~al.}(2015)\citenamefont{{Udalski},
  {Szyma{\'n}ski}, and {Szyma{\'n}ski}}}]{2015AcA....65....1U}
\bibinfo{author}{\bibfnamefont{A.}~\bibnamefont{{Udalski}}},
  \bibinfo{author}{\bibfnamefont{M.~K.} \bibnamefont{{Szyma{\'n}ski}}},
  \bibnamefont{and}
  \bibinfo{author}{\bibfnamefont{G.}~\bibnamefont{{Szyma{\'n}ski}}},
  \bibinfo{journal}{Acta~Astronomica} \textbf{\bibinfo{volume}{65}},
  \bibinfo{pages}{1} (\bibinfo{year}{2015}), \eprint{1504.05966}.

\bibitem[{\citenamefont{{Mr{\'o}z} et~al.}(2017)\citenamefont{{Mr{\'o}z},
  {Udalski}, {Skowron}, {Poleski}, {Koz{\l}owski}, {Szyma{\'n}ski},
  {Soszy{\'n}ski}, {Wyrzykowski}, {Pietrukowicz}, {Ulaczyk}
  et~al.}}]{2017Natur.548..183M}
\bibinfo{author}{\bibfnamefont{P.}~\bibnamefont{{Mr{\'o}z}}},
  \bibinfo{author}{\bibfnamefont{A.}~\bibnamefont{{Udalski}}},
  \bibinfo{author}{\bibfnamefont{J.}~\bibnamefont{{Skowron}}},
  \bibinfo{author}{\bibfnamefont{R.}~\bibnamefont{{Poleski}}},
  \bibinfo{author}{\bibfnamefont{S.}~\bibnamefont{{Koz{\l}owski}}},
  \bibinfo{author}{\bibfnamefont{M.~K.} \bibnamefont{{Szyma{\'n}ski}}},
  \bibinfo{author}{\bibfnamefont{I.}~\bibnamefont{{Soszy{\'n}ski}}},
  \bibinfo{author}{\bibfnamefont{{\L}.}~\bibnamefont{{Wyrzykowski}}},
  \bibinfo{author}{\bibfnamefont{P.}~\bibnamefont{{Pietrukowicz}}},
  \bibinfo{author}{\bibfnamefont{K.}~\bibnamefont{{Ulaczyk}}},
  \bibnamefont{et~al.}, \bibinfo{journal}{\nat} \textbf{\bibinfo{volume}{548}},
  \bibinfo{pages}{183} (\bibinfo{year}{2017}), \eprint{1707.07634}.

\bibitem[{\citenamefont{{Sumi} et~al.}(2003)\citenamefont{{Sumi}, {Abe},
  {Bond}, {Dodd}, {Hearnshaw}, {Honda}, {Honma}, {Kan-ya}, {Kilmartin},
  {Masuda} et~al.}}]{Sumietal:03}
\bibinfo{author}{\bibfnamefont{T.}~\bibnamefont{{Sumi}}},
  \bibinfo{author}{\bibfnamefont{F.}~\bibnamefont{{Abe}}},
  \bibinfo{author}{\bibfnamefont{I.~A.} \bibnamefont{{Bond}}},
  \bibinfo{author}{\bibfnamefont{R.~J.} \bibnamefont{{Dodd}}},
  \bibinfo{author}{\bibfnamefont{J.~B.} \bibnamefont{{Hearnshaw}}},
  \bibinfo{author}{\bibfnamefont{M.}~\bibnamefont{{Honda}}},
  \bibinfo{author}{\bibfnamefont{M.}~\bibnamefont{{Honma}}},
  \bibinfo{author}{\bibfnamefont{Y.}~\bibnamefont{{Kan-ya}}},
  \bibinfo{author}{\bibfnamefont{P.~M.} \bibnamefont{{Kilmartin}}},
  \bibinfo{author}{\bibfnamefont{K.}~\bibnamefont{{Masuda}}},
  \bibnamefont{et~al.}, \bibinfo{journal}{\apj} \textbf{\bibinfo{volume}{591}},
  \bibinfo{pages}{204} (\bibinfo{year}{2003}), \eprint{astro-ph/0207604}.

\bibitem[{\citenamefont{{Sumi} et~al.}(2011)\citenamefont{{Sumi}, {Kamiya},
  {Bennett}, {Bond}, {Abe}, {Botzler}, {Fukui}, {Furusawa}, {Hearnshaw}, {Itow}
  et~al.}}]{Sumietal:11}
\bibinfo{author}{\bibfnamefont{T.}~\bibnamefont{{Sumi}}},
  \bibinfo{author}{\bibfnamefont{K.}~\bibnamefont{{Kamiya}}},
  \bibinfo{author}{\bibfnamefont{D.~P.} \bibnamefont{{Bennett}}},
  \bibinfo{author}{\bibfnamefont{I.~A.} \bibnamefont{{Bond}}},
  \bibinfo{author}{\bibfnamefont{F.}~\bibnamefont{{Abe}}},
  \bibinfo{author}{\bibfnamefont{C.~S.} \bibnamefont{{Botzler}}},
  \bibinfo{author}{\bibfnamefont{A.}~\bibnamefont{{Fukui}}},
  \bibinfo{author}{\bibfnamefont{K.}~\bibnamefont{{Furusawa}}},
  \bibinfo{author}{\bibfnamefont{J.~B.} \bibnamefont{{Hearnshaw}}},
  \bibinfo{author}{\bibfnamefont{Y.}~\bibnamefont{{Itow}}},
  \bibnamefont{et~al.}, \bibinfo{journal}{\nat} \textbf{\bibinfo{volume}{473}},
  \bibinfo{pages}{349} (\bibinfo{year}{2011}), \eprint{1105.3544}.

\bibitem[{\citenamefont{{Mroz} et~al.}(2018)\citenamefont{{Mroz}, {Udalski},
  {Bennett}, {Ryu}, {Sumi}, {Shvartzvald}, {Skowron}, {Poleski},
  {Pietrukowicz}, {Kozlowski} et~al.}}]{2018arXiv181100441M}
\bibinfo{author}{\bibfnamefont{P.}~\bibnamefont{{Mroz}}},
  \bibinfo{author}{\bibfnamefont{A.}~\bibnamefont{{Udalski}}},
  \bibinfo{author}{\bibfnamefont{D.~P.} \bibnamefont{{Bennett}}},
  \bibinfo{author}{\bibfnamefont{Y.-H.} \bibnamefont{{Ryu}}},
  \bibinfo{author}{\bibfnamefont{T.}~\bibnamefont{{Sumi}}},
  \bibinfo{author}{\bibfnamefont{Y.}~\bibnamefont{{Shvartzvald}}},
  \bibinfo{author}{\bibfnamefont{J.}~\bibnamefont{{Skowron}}},
  \bibinfo{author}{\bibfnamefont{R.}~\bibnamefont{{Poleski}}},
  \bibinfo{author}{\bibfnamefont{P.}~\bibnamefont{{Pietrukowicz}}},
  \bibinfo{author}{\bibfnamefont{S.}~\bibnamefont{{Kozlowski}}},
  \bibnamefont{et~al.}, \bibinfo{journal}{arXiv e-prints}
  (\bibinfo{year}{2018}), \eprint{1811.00441}.

\bibitem[{\citenamefont{{Han} and {Gould}}(1995)}]{1995ApJ...447...53H}
\bibinfo{author}{\bibfnamefont{C.}~\bibnamefont{{Han}}} \bibnamefont{and}
  \bibinfo{author}{\bibfnamefont{A.}~\bibnamefont{{Gould}}},
  \bibinfo{journal}{\apj} \textbf{\bibinfo{volume}{447}}, \bibinfo{pages}{53}
  (\bibinfo{year}{1995}), \eprint{astro-ph/9409036}.

\bibitem[{\citenamefont{{Han} and {Gould}}(1996)}]{1996ApJ...467..540H}
\bibinfo{author}{\bibfnamefont{C.}~\bibnamefont{{Han}}} \bibnamefont{and}
  \bibinfo{author}{\bibfnamefont{A.}~\bibnamefont{{Gould}}},
  \bibinfo{journal}{\apj} \textbf{\bibinfo{volume}{467}}, \bibinfo{pages}{540}
  (\bibinfo{year}{1996}), \eprint{astro-ph/9504078}.

\bibitem[{\citenamefont{{Meylan} et~al.}(2006)\citenamefont{{Meylan}, {Jetzer},
  {North}, {Schneider}, {Kochanek}, and {Wambsganss}}}]{2006glsw.conf.....M}
\bibinfo{editor}{\bibfnamefont{G.}~\bibnamefont{{Meylan}}},
  \bibinfo{editor}{\bibfnamefont{P.}~\bibnamefont{{Jetzer}}},
  \bibinfo{editor}{\bibfnamefont{P.}~\bibnamefont{{North}}},
  \bibinfo{editor}{\bibfnamefont{P.}~\bibnamefont{{Schneider}}},
  \bibinfo{editor}{\bibfnamefont{C.~S.} \bibnamefont{{Kochanek}}},
  \bibnamefont{and}
  \bibinfo{editor}{\bibfnamefont{J.}~\bibnamefont{{Wambsganss}}}, eds.,
  \emph{\bibinfo{title}{{Gravitational Lensing: Strong, Weak and Micro}}}
  (\bibinfo{year}{2006}), \eprint{astro-ph/0407232}.

\bibitem[{\citenamefont{{Dodelson}}(2017)}]{2017grle.book.....D}
\bibinfo{author}{\bibfnamefont{S.}~\bibnamefont{{Dodelson}}},
  \emph{\bibinfo{title}{{Gravitational Lensing}}} (\bibinfo{year}{2017}).

\bibitem[{\citenamefont{{Kent}}(1992)}]{1992ApJ...387..181K}
\bibinfo{author}{\bibfnamefont{S.~M.} \bibnamefont{{Kent}}},
  \bibinfo{journal}{\apj} \textbf{\bibinfo{volume}{387}}, \bibinfo{pages}{181}
  (\bibinfo{year}{1992}).

\bibitem[{\citenamefont{{Bahcall}}(1986)}]{1986ARA&A..24..577B}
\bibinfo{author}{\bibfnamefont{J.~N.} \bibnamefont{{Bahcall}}},
  \bibinfo{journal}{\araa} \textbf{\bibinfo{volume}{24}}, \bibinfo{pages}{577}
  (\bibinfo{year}{1986}).

\bibitem[{\citenamefont{{Navarro} et~al.}(1997)\citenamefont{{Navarro},
  {Frenk}, and {White}}}]{NFW97}
\bibinfo{author}{\bibfnamefont{J.~F.} \bibnamefont{{Navarro}}},
  \bibinfo{author}{\bibfnamefont{C.~S.} \bibnamefont{{Frenk}}},
  \bibnamefont{and} \bibinfo{author}{\bibfnamefont{S.~D.~M.}
  \bibnamefont{{White}}}, \bibinfo{journal}{\apj}
  \textbf{\bibinfo{volume}{490}}, \bibinfo{pages}{493} (\bibinfo{year}{1997}),
  \eprint{astro-ph/9611107}.

\bibitem[{\citenamefont{{Klypin} et~al.}(2002)\citenamefont{{Klypin}, {Zhao},
  and {Somerville}}}]{Klypinetal:02}
\bibinfo{author}{\bibfnamefont{A.}~\bibnamefont{{Klypin}}},
  \bibinfo{author}{\bibfnamefont{H.}~\bibnamefont{{Zhao}}}, \bibnamefont{and}
  \bibinfo{author}{\bibfnamefont{R.~S.} \bibnamefont{{Somerville}}},
  \bibinfo{journal}{\apj} \textbf{\bibinfo{volume}{573}}, \bibinfo{pages}{597}
  (\bibinfo{year}{2002}), \eprint{astro-ph/0110390}.

\bibitem[{\citenamefont{{Callingham} et~al.}(2018)\citenamefont{{Callingham},
  {Cautun}, {Deason}, {Frenk}, {Wang}, {G{\'o}mez}, {Grand}, {Marinacci}, and
  {Pakmor}}}]{2018arXiv180810456C}
\bibinfo{author}{\bibfnamefont{T.}~\bibnamefont{{Callingham}}},
  \bibinfo{author}{\bibfnamefont{M.}~\bibnamefont{{Cautun}}},
  \bibinfo{author}{\bibfnamefont{A.~J.} \bibnamefont{{Deason}}},
  \bibinfo{author}{\bibfnamefont{C.~S.} \bibnamefont{{Frenk}}},
  \bibinfo{author}{\bibfnamefont{W.}~\bibnamefont{{Wang}}},
  \bibinfo{author}{\bibfnamefont{F.~A.} \bibnamefont{{G{\'o}mez}}},
  \bibinfo{author}{\bibfnamefont{R.~J.~J.} \bibnamefont{{Grand}}},
  \bibinfo{author}{\bibfnamefont{F.}~\bibnamefont{{Marinacci}}},
  \bibnamefont{and} \bibinfo{author}{\bibfnamefont{R.}~\bibnamefont{{Pakmor}}},
  \bibinfo{journal}{arXiv e-prints} \bibinfo{eid}{arXiv:1808.10456}
  (\bibinfo{year}{2018}), \eprint{1808.10456}.

\bibitem[{\citenamefont{{Vogelsberger}
  et~al.}(2009)\citenamefont{{Vogelsberger}, {Helmi}, {Springel}, {White},
  {Wang}, {Frenk}, {Jenkins}, {Ludlow}, and {Navarro}}}]{2009MNRAS.395..797V}
\bibinfo{author}{\bibfnamefont{M.}~\bibnamefont{{Vogelsberger}}},
  \bibinfo{author}{\bibfnamefont{A.}~\bibnamefont{{Helmi}}},
  \bibinfo{author}{\bibfnamefont{V.}~\bibnamefont{{Springel}}},
  \bibinfo{author}{\bibfnamefont{S.~D.~M.} \bibnamefont{{White}}},
  \bibinfo{author}{\bibfnamefont{J.}~\bibnamefont{{Wang}}},
  \bibinfo{author}{\bibfnamefont{C.~S.} \bibnamefont{{Frenk}}},
  \bibinfo{author}{\bibfnamefont{A.}~\bibnamefont{{Jenkins}}},
  \bibinfo{author}{\bibfnamefont{A.}~\bibnamefont{{Ludlow}}}, \bibnamefont{and}
  \bibinfo{author}{\bibfnamefont{J.~F.} \bibnamefont{{Navarro}}},
  \bibinfo{journal}{\mnras} \textbf{\bibinfo{volume}{395}},
  \bibinfo{pages}{797} (\bibinfo{year}{2009}), \eprint{0812.0362}.

\bibitem[{\citenamefont{{Gould}}(2000)}]{2000ApJ...535..928G}
\bibinfo{author}{\bibfnamefont{A.}~\bibnamefont{{Gould}}},
  \bibinfo{journal}{\apj} \textbf{\bibinfo{volume}{535}}, \bibinfo{pages}{928}
  (\bibinfo{year}{2000}), \eprint{astro-ph/9906472}.

\bibitem[{\citenamefont{{Kroupa}}(2001)}]{2001MNRAS.322..231K}
\bibinfo{author}{\bibfnamefont{P.}~\bibnamefont{{Kroupa}}},
  \bibinfo{journal}{\mnras} \textbf{\bibinfo{volume}{322}},
  \bibinfo{pages}{231} (\bibinfo{year}{2001}), \eprint{astro-ph/0009005}.

\bibitem[{\citenamefont{{Zoccali} et~al.}(2000)\citenamefont{{Zoccali},
  {Cassisi}, {Frogel}, {Gould}, {Ortolani}, {Renzini}, {Rich}, and
  {Stephens}}}]{2000ApJ...530..418Z}
\bibinfo{author}{\bibfnamefont{M.}~\bibnamefont{{Zoccali}}},
  \bibinfo{author}{\bibfnamefont{S.}~\bibnamefont{{Cassisi}}},
  \bibinfo{author}{\bibfnamefont{J.~A.} \bibnamefont{{Frogel}}},
  \bibinfo{author}{\bibfnamefont{A.}~\bibnamefont{{Gould}}},
  \bibinfo{author}{\bibfnamefont{S.}~\bibnamefont{{Ortolani}}},
  \bibinfo{author}{\bibfnamefont{A.}~\bibnamefont{{Renzini}}},
  \bibinfo{author}{\bibfnamefont{R.~M.} \bibnamefont{{Rich}}},
  \bibnamefont{and} \bibinfo{author}{\bibfnamefont{A.~W.}
  \bibnamefont{{Stephens}}}, \bibinfo{journal}{\apj}
  \textbf{\bibinfo{volume}{530}}, \bibinfo{pages}{418} (\bibinfo{year}{2000}),
  \eprint{astro-ph/9906452}.

\bibitem[{\citenamefont{{Burrows} et~al.}(2001)\citenamefont{{Burrows},
  {Hubbard}, {Lunine}, and {Liebert}}}]{2001RvMP...73..719B}
\bibinfo{author}{\bibfnamefont{A.}~\bibnamefont{{Burrows}}},
  \bibinfo{author}{\bibfnamefont{W.~B.} \bibnamefont{{Hubbard}}},
  \bibinfo{author}{\bibfnamefont{J.~I.} \bibnamefont{{Lunine}}},
  \bibnamefont{and}
  \bibinfo{author}{\bibfnamefont{J.}~\bibnamefont{{Liebert}}},
  \bibinfo{journal}{Reviews of Modern Physics} \textbf{\bibinfo{volume}{73}},
  \bibinfo{pages}{719} (\bibinfo{year}{2001}), \eprint{astro-ph/0103383}.

\bibitem[{\citenamefont{{Inomata} et~al.}(2018)\citenamefont{{Inomata},
  {Kawasaki}, {Mukaida}, and {Yanagida}}}]{2018PhRvD..97d3514I}
\bibinfo{author}{\bibfnamefont{K.}~\bibnamefont{{Inomata}}},
  \bibinfo{author}{\bibfnamefont{M.}~\bibnamefont{{Kawasaki}}},
  \bibinfo{author}{\bibfnamefont{K.}~\bibnamefont{{Mukaida}}},
  \bibnamefont{and} \bibinfo{author}{\bibfnamefont{T.~T.}
  \bibnamefont{{Yanagida}}}, \bibinfo{journal}{\prd}
  \textbf{\bibinfo{volume}{97}}, \bibinfo{eid}{043514} (\bibinfo{year}{2018}),
  \eprint{1711.06129}.

\bibitem[{\citenamefont{{Inomata}
  et~al.}(2017{\natexlab{b}})\citenamefont{{Inomata}, {Kawasaki}, {Mukaida},
  {Tada}, and {Yanagida}}}]{Inomataetal:17}
\bibinfo{author}{\bibfnamefont{K.}~\bibnamefont{{Inomata}}},
  \bibinfo{author}{\bibfnamefont{M.}~\bibnamefont{{Kawasaki}}},
  \bibinfo{author}{\bibfnamefont{K.}~\bibnamefont{{Mukaida}}},
  \bibinfo{author}{\bibfnamefont{Y.}~\bibnamefont{{Tada}}}, \bibnamefont{and}
  \bibinfo{author}{\bibfnamefont{T.~T.} \bibnamefont{{Yanagida}}},
  \bibinfo{journal}{\prd} \textbf{\bibinfo{volume}{96}}, \bibinfo{eid}{043504}
  (\bibinfo{year}{2017}{\natexlab{b}}), \eprint{1701.02544}.

\end{thebibliography}

\end{document}